# Extended and revised analysis of singly ionized tin: Sn II

**K Haris[1], A Kramida[2] and A Tauheed[1]**

[1] Department of Physics, Aligarh Muslim University, Aligarh 202002, India
[2] National Institute of Standards and Technology, Gaithersburg, MD 20899, USA

E-mail: alexander.kramida@nist.gov

**Abstract.** The electronic structure of singly ionized tin (Sn II) is partly a one-electron and partly a three-electron system with ground configuration $5s^25p$. The excited configurations are of the type $5s^2n\ell$ in the one-electron part, and $5s5p^2$, $5p^3$ and $5s5pn\ell$ ($n\ell$ = 6s, 5d) in the three-electron system with quartet and doublet levels. The spectrum analyzed in this work was recorded on a 3 m normal incidence vacuum spectrograph of the Antigonish laboratory (Canada) in the wavelength region 300 Å to 2080 Å using a triggered spark source. The existing interpretation of the one-electron level system was confirmed in this paper, while the $^2S_{1/2}$ level of the $5s5p^2$ configuration has been revised. The analysis has been extended to include new configurations $5p^3$, 5s5p5d and 5s5p6s with the aid of superposition-of-configurations Hartree–Fock calculations with relativistic corrections. The ionization potential obtained from the *n*g series was found to be 118023.7(5) cm$^{-1}$ (14.63307(6) eV). We give a complete set of critically evaluated data on energy levels, observed wavelengths and transition probabilities of Sn II in the range 888 Å to 10740 Å involving excitation of the $n$ = 5 electrons.

(Some figures may appear in colour only in the online journal)

## 1. Introduction

Accurate data on the spectrum of singly ionized tin are needed in different fields of scientific research and industry. Such data are useful for astrophysical observations, development of various light sources, and for plasma diagnostics in fusion power plants. The astrophysical importance of tin has increased since gas-phase tin was first detected by Hobbs et al. [1] in the spectra acquired with the Goddard High Resolution Spectrograph on board the Hubble Space Telescope. They observed the absorption line of Sn II at 1400.45 Å from various interstellar sources. Later, the same line was observed in diffuse interstellar clouds by Sofia et al. [2]. They discovered that the gas-phase abundance of Sn in the interstellar medium (ISM) appears to be supersolar, which further substantiates the slow neutron capture (s-process) enrichment believed to be a major contributor to the nuclearsynthesis of elements beyond the iron peak in the ISM. In erosion probing of vessel wall tiles of future fusion power plants, such as ITER, spectroscopic data on tin may play a major diagnostic role [3].

Singly ionized tin (Sn II) is the second member of the In I isoelectronic sequence with the ground configuration $4d^{10}5s^25p$ consisting of the ground level $^2P°_{1/2}$ and first excited level $^2P°_{3/2}$. The currently available spectroscopic information on Sn II compiled in Moore's Atomic Energy Levels (AEL) compilation [4] and listed in the Atomic Spectra Database (ASD) [5] of the National Institute of Standards and Technology (NIST) is based on an unpublished work of Shenstone. Prior to AEL, extensive work in this spectrum was carried out by McCormick and Sawyer [6], who revised the earlier findings of Green and Loring [7], Narayan and Rao [8], and Lang [9]. Shenstone in his work quoted in AEL re-investigated this spectrum in the wavelength range of 600 Å to 2500 Å and extended the analysis to include the 5s5p5d and 5s5p6s configurations. Shenstone revised some energy levels of Sn II and improved the accuracy of the earlier reported energy level values on the basis of his observations. Some spectral lines of Sn II were also reported by Brill [10] in his doctoral thesis and by Wu [11] in his master thesis. Apart from spectroscopy of valence-shell electrons, spectral studies of the 4d-core excitation of Sn II in the extreme ultraviolet (EUV) region with the dual laser plasma (DLP) method were made by Lysaght et al. [12] and by Duffy et al. [13].

Despite those extensive studies, the currently available data are still inadequate, since there are considerable anomalies in energy level values and line assignments. Many of the energy levels given in AEL [4] are not supported by any published line lists. The lines determining these energy levels have to be re-discovered.

There are many theoretical studies on radiative lifetimes, transition rates, and oscillator strengths of Sn II. Among them, the most accurate and reliable work was done by Oliver and Hibbert [14]. Accurate lifetime measurements, and thereby *f*-value determinations, were made by Schectman et al. [15] who improved the earlier work of Andersen and Lindgård [16]. Data from the latter work were used by Sofia et al. [2] to derive the abundance of tin.

In the present work, our motivation is to provide a comprehensive spectroscopic analysis of singly ionized tin on the basis of tin spectra taken by us, the tin spectral line list given by Wu [11], the Sn II spectral classification by McCormick and Sawyer [6], and lines reported by Brill [10]. All previously reported energy levels of this spectrum are subjected to a critical investigation. One of our goals is to resolve the anomaly existing in the $5s5p^2$ $^2S_{1/2}$ and $^2P_{1/2}$ level values. Excitation from the $5s5p^2$ configuration to 5s5p(5d+6s) and $5p^3$ is studied extensively in this work. Although some of the levels of these highly-excited configurations were tentatively identified by Shenstone and listed in AEL [4], Shenstone's analysis was incomplete in many respects. Some of the level values were given with question marks, and some had uncertain $J$ values and/or designations. A very recent study carried out by Alonso-Medina et al. [17] using laser-produced plasma of a Sn/Pb target reproduced some of the levels reported in AEL [4], but the majority of their suggested 5s5p5d level assignments and line classifications are made on the basis of a physically inadequate theoretical atomic model. We attempt to resolve all these ambiguities in the present analysis. Interestingly, many of the 5s5p5d and $5p^3$ levels are located above the first ionization limit. Therefore, only those levels that have autoionization rates comparable to or smaller than the radiative decay rate were observed via their corresponding photon decay channel.

Although, as noted above, some studies of the 4d core-excited spectrum of Sn II have been published [12, 13], we restrict the scope of this paper to excitations of the $n$ = 5 electrons.

## 2. Experimental details

The tin ions/atoms were excited by means of a triggered spark source, which consists of a 14.5 $\mu$F fast-charging low-inductance capacitor, chargeable up to 20 kV, and a trigger module to initiate the discharge. Either pure electrodes made of tin, or tin samples inserted into a cavity in aluminum electrodes were used. The tin spectrum was recorded at St. Francis Xavier University, Antigonish (Canada) using a 3 m normal incidence vacuum ultraviolet (VUV) spectrograph in the (300–2080) Å wavelength region. A holographic osmium-coated grating with 2400 lines/mm was used to obtain reciprocal linear dispersion of about 1.4 Å/mm in the aforementioned region in the first order of diffraction. At least four or five different tracks of spectrum were photographed on a Kodak SWR[1] (short-wave radiation) plates with varied experimental conditions, such as electric current and voltage. The purpose of the different exposures was to distinguish the lines of Sn II from other ionization species. This was achieved by inserting a low, medium, or high inductance in series with the spark circuit or by varying the charging potential within the limits of 2 kV and 6 kV. The inductances were made of copper wire, 2 mm in diameter, wound on a cylinder of diameter 24 cm in turns separated by about 4 mm. A low inductance coil had 8 or 9 turns of wire, a medium one had 25 turns, and the high inductance one had 40 to 50 turns. The optimal conditions for observing the Sn II spectrum were found to be at 2 kV without an additional inductance or at 4 kV with a medium inductance. Relative positions of spectral lines on the plates were measured using a Zeiss Abbe[1] comparator at the Aligarh University (India). For their wavelength reduction, the known impurity lines of C, O, Al, and Si [18] were used as internal standards to obtain a second- or third-degree polynomial fit with a mean deviation of 0.005 Å or less. This value represents the wavelength uncertainty of our measurements for sharp and unperturbed lines. A more detailed discussion of uncertainties is given in the next section.

[1] Commercial products are identified in this paper for adequate specification of the experimental procedure. This identification does not imply recommendation or endorsement by NIST.

## 3. Measurement uncertainties

The general estimate of uncertainty given in the previous section, 0.005 Å for strong unperturbed lines, is insufficient for deriving accurate energy level values from the observed wavelengths. For that purpose, it is necessary to estimate the uncertainty for each individual line. We examined all our observed lines and assigned somewhat greater uncertainties to blended lines and those that appear to be broadened and/or asymmetric. The largest uncertainty, 0.010 Å, was assigned to blended and hazy lines. Many of the latter category lines appear to be broadened by autoionization of the upper level. The values of uncertainty assigned to each line can be found in table 1. All uncertainties reported in the present work are meant to be on the level of one standard deviation.

Many of the known classified lines of Sn II were observed by other researchers [6, 10, 11] outside the wavelength range studied in the present work. Thus, to obtain optimized level values, wavelength values and uncertainties reported by other observers have to be evaluated.

The most valuable of the previously reported measurements are those of Brill [10]. He reported 42 wavelengths of Sn II between 2150 Å and 10740 Å, with uncertainties estimated individually for each line. For 39 of these lines, the measurements were made interferometrically, and their uncertainties vary between 0.0006 Å and 0.006 Å. Three weak lines were measured with a grating spectrograph; their uncertainties are between 0.06 Å and 0.09 Å.

Two other large sets of wavelengths were taken from Wu's thesis [11] and from McCormick and Sawyer [6]. Wu photographed the tin spectra in the region between 350 Å and 9000 Å using an electrodeless discharge. A condensed spark in helium with a 3 m normal incidence vacuum grating spectrograph and a prism spectrograph were used in the regions below and above 2400 Å, respectively. Although all wavelengths in Wu's line list were given with three digits after the decimal point (in angstroms), the wavelength uncertainty varied greatly depending on the wavelength region and on the spectrographs used. To assess the uncertainties of Wu's wavelength measurements, we compared his reported wavelengths with more accurate ones taken from ASD [5] for Sn I and Sn III and with the Sn II Ritz wavelengths (see section 4) determined mainly by our measurements in the VUV and by Brill's measurements [10] in the air region. This comparison, shown in figures 1a, 1b, and 1c, revealed significant systematic shifts. These shifts vary smoothly with wavelength between +0.019 Å near 900 Å and −0.25 Å near 8300 Å. After removing these shifts, the measurement uncertainties of the corrected wavelengths were estimated from their average deviations from reference values. In the region below 2400 Å, where the grating spectrograph was used, the estimated wavelength uncertainty is almost constant, about 0.019 Å. In the region 2400 Å to 3050 Å, where the quartz prism spectrograph was used, the uncertainties are about 0.024 Å on average. However, uncertainties of Wu's prism spectra are better described by a constant uncertainty in wavenumber, about 0.3 $cm^{-1}$ for this wavelength region. This implies a gradual increase of uncertainties from 0.019 Å at 2400 Å to 0.03 Å at 3050 Å. Above this wavelength, as figure 1c shows, uncertainties increase abruptly to 1.7 $cm^{-1}$, corresponding to 0.16 Å at 3100 Å and 1.2 Å at 8300 Å.

McCormick and Sawyer [6] excited the Sn II spectrum in a hollow cathode discharge in helium and photographed it in a similarly wide wavelength range from 800 Å to 10 000 Å. In the region below 2200 Å, they used a 1 m vacuum grating spectrograph. The region from 2200 Å to 2700 Å was photographed with a quartz prism spectrograph. Above 2700 Å, two other prism spectrographs were used. Since these authors reported only the Sn II wavelengths, the only means of assessment of their uncertainties was a comparison with more accurate Ritz wavelengths. For this comparison, we used the Ritz wavelengths from our preliminary level optimization (see section 4) that were mainly determined by our VUV measurements and those of Brill [10] in the region above 2150 Å. Figure 1d shows these deviations, scaled in such a way that their scatter has similar magnitudes throughout the wavelength range covered by the figure.

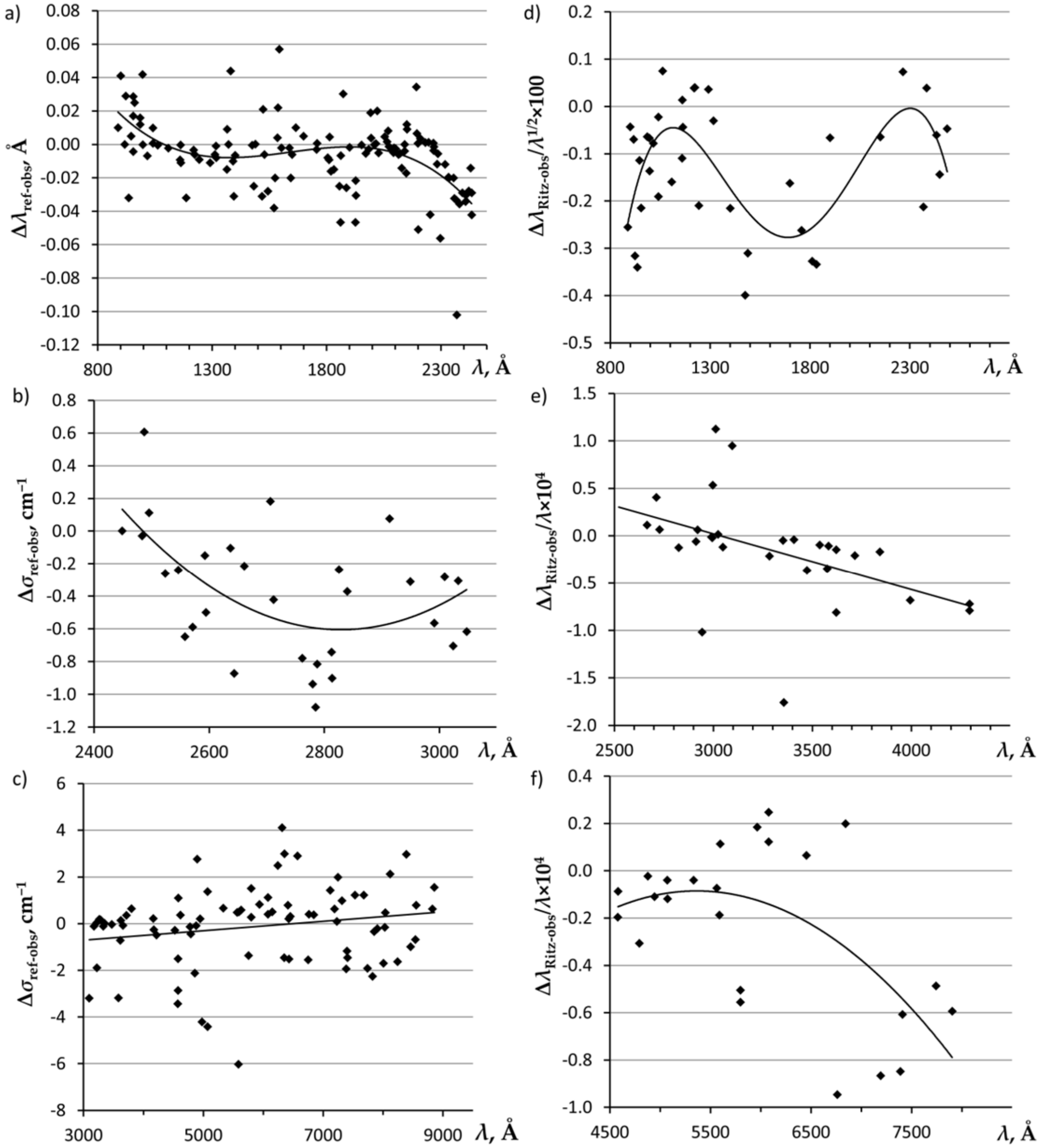

Figure 1. Differences between observed and reference wavelengths $\lambda$ or wave numbers $\sigma$ for the measurements of Wu [11] (a, b, c) and McCormick and Sawyer [6] (d, e, f). The solid lines are linear or polynomial fits determining the systematic corrections to the original measurements.

As with the work of Wu, measurements of McCormick and Sawyer [6] appear to have significant systematic shifts smoothly varying with wavelength, from −0.11 Å at 1700 Å to zero at 2300 Å. At longer wavelengths, as figures 1e and 1f show, statistical uncertainties appear to be almost constant, if they are scaled by dividing them by wavelength. Systematic shifts are significant in these regions as well, varying from +0.08 Å at about 3000 Å to −0.6 Å at 8000 Å.

All identified lines of Sn II are collected in table 1 with the adopted wavelengths and their uncertainties. In total, there are about 200 lines, of which 70 were measured in the present work, 42 are from Brill [10], 27 are from McCormick and Sawyer [6], and 63 are from Wu [11]. Among Wu's lines [11], 12 were classified as Sn II transitions by us.

Table 1. Classified lines in Sn II

| $I_{obs}$ [a] arb. u. | Char. [b] | $\lambda_{obs}$ [c] Å | $\sigma_{obs}$ cm$^{-1}$ | $\lambda_{Ritz}$ [d] Å | $\delta\lambda_{O-Ritz}$ [e] Å | Classification | | | | $E_{low}$ cm$^{-1}$ | $E_{upp}$ cm$^{-1}$ | $A$ [f] s$^{-1}$ | Acc. [g] | Line Ref. [h] | TP Ref. [h] |
|---|---|---|---|---|---|---|---|---|---|---|---|---|---|---|---|
| 1400 | | 888.313(19) | 112572.9 | 888.304(4) | 0.009 | $5s^25p$ | $^2P^\circ_{1/2}$ | $5s^211d$ | $^2D_{3/2}$ | 0.00 | 112574.1 | | | Wu | |
| 3400 | | 899.884(19) | 111125.4 | 899.907(10) | −0.023 | $5s^25p$ | $^2P^\circ_{1/2}$ | $5s^210d$ | $^2D_{3/2}$ | 0.00 | 111122.6 | | | Wu | |
| 7300 | | 917.378(5) | 109006.3 | 917.379(3) | −0.001 | $5s^25p$ | $^2P^\circ_{1/2}$ | $5s^29d$ | $^2D_{3/2}$ | 0.00 | 109006.2 | 9.+7 | E | TW | TW |
| 3500 | | 922.856(19) | 108359.3 | 922.870(3) | −0.014 | $5s^25p$ | $^2P^\circ_{1/2}$ | $5s^210s$ | $^2S_{1/2}$ | 0.00 | 108357.6 | 2.4+7 | E | Wu | TW |
| 3700 | | 923.01(4) | 108341 | 922.974(10) | 0.04 | $5s^25p$ | $^2P^\circ_{3/2}$ | $5s^211d$ | $^2D_{5/2}$ | 4251.505 | 112596.9 | | | MS | |
| 5800 | | 935.571(19) | 106886.6 | 935.526(10) | 0.045 | $5s^25p$ | $^2P^\circ_{3/2}$ | $5s^210d$ | $^2D_{5/2}$ | 4251.505 | 111143.3 | | | Wu | |
| 14000 | | 945.801(5) | 105730.5 | 945.795(3) | 0.006 | $5s^25p$ | $^2P^\circ_{1/2}$ | $5s^28d$ | $^2D_{3/2}$ | 0.00 | 105731.2 | 1.5+8 | E | TW | TW |
| 12000 | | 954.445(5) | 104772.9 | 954.4337(14) | 0.011 | $5s^25p$ | $^2P^\circ_{3/2}$ | $5s^29d$ | $^2D_{5/2}$ | 4251.505 | 109025.68 | 9.+7 | E | TW | TW |
| 6200 | | 954.612(5) | 104754.6 | 954.611(3) | 0.001 | $5s^25p$ | $^2P^\circ_{3/2}$ | $5s^29d$ | $^2D_{3/2}$ | 4251.505 | 109006.2 | 1.8+7 | E | TW | TW |
| 6900 | | 955.301(5) | 104679.0 | 955.3066(10) | −0.006 | $5s^25p$ | $^2P^\circ_{1/2}$ | $5s^29s$ | $^2S_{1/2}$ | 0.00 | 104678.43 | 3.9+7 | E | TW | TW |
| 2400 | | 960.545(19) | 104107.6 | 960.559(4) | −0.014 | $5s^25p$ | $^2P^\circ_{3/2}$ | $5s^210s$ | $^2S_{1/2}$ | 4251.505 | 108357.6 | 4.+7 | E | Wu | TW |
| 18000 | | 985.101(5) | 101512.4 | 985.1099(22) | −0.009 | $5s^25p$ | $^2P^\circ_{3/2}$ | $5s^28d$ | $^2D_{5/2}$ | 4251.505 | 105763.02 | 1.6+8 | E | TW | TW |
| 10000 | | 985.411(19) | 101480.5 | 985.419(3) | −0.008 | $5s^25p$ | $^2P^\circ_{3/2}$ | $5s^28d$ | $^2D_{3/2}$ | 4251.505 | 105731.2 | 3.0+7 | E | Wu | TW |
| 8600 | | 995.743(5) | 100427.5 | 995.7489(10) | −0.006 | $5s^25p$ | $^2P^\circ_{3/2}$ | $5s^29s$ | $^2S_{1/2}$ | 4251.505 | 104678.43 | 7.+7 | E | TW | TW |
| 19000 | | 997.157(5) | 100285.1 | 997.1668(5) | −0.010 | $5s^25p$ | $^2P^\circ_{1/2}$ | $5s^27d$ | $^2D_{3/2}$ | 0.00 | 100284.125 | 2.8+8 | D+ | TW | TW |
| 16000 | | 1016.240(5) | 98402.0 | 1016.2351(5) | 0.005 | $5s^25p$ | $^2P^\circ_{1/2}$ | $5s^28s$ | $^2S_{1/2}$ | 0.00 | 98402.425 | 7.+7 | E | TW | TW |
| 24000 | | 1040.722(5) | 96087.1 | 1040.71858(18) | 0.003 | $5s^25p$ | $^2P^\circ_{3/2}$ | $5s^27d$ | $^2D_{5/2}$ | 4251.505 | 100338.960 | 3.0+8 | D+ | TW | TW |
| 16000 | | 1041.315(5) | 96032.4 | 1041.31284(19) | 0.002 | $5s^25p$ | $^2P^\circ_{3/2}$ | $5s^27d$ | $^2D_{3/2}$ | 4251.505 | 100284.125 | 6.+7 | E | TW | TW |
| 25000 | | 1062.126(5) | 94150.8 | 1062.12451(17) | 0.001 | $5s^25p$ | $^2P^\circ_{3/2}$ | $5s^28s$ | $^2S_{1/2}$ | 4251.505 | 98402.425 | 1.3+8 | E | TW | TW |
| 43000 | | 1108.130(7) | 90242.1 | 1108.1368(5) | −0.007 | $5s^25p$ | $^2P^\circ_{1/2}$ | $5s^26d$ | $^2D_{3/2}$ | 0.00 | 90241.568 | 4.7+8 | B+ | TW | OH10 |
| 53000 | | 1159.013(7) | 86280.3 | 1159.0127(6) | 0.000 | $5s^25p$ | $^2P^\circ_{1/2}$ | $5s^27s$ | $^2S_{1/2}$ | 0.00 | 86280.332 | 1.01+8 | C+ | TW | OH10 |
| 74000 | | 1161.427(7) | 86101.0 | 1161.43475(20) | −0.008 | $5s^25p$ | $^2P^\circ_{3/2}$ | $5s^26d$ | $^2D_{5/2}$ | 4251.505 | 90351.908 | 5.5+8 | B+ | TW | OH10 |
| 56000 | | 1162.920(7) | 85990.4 | 1162.92507(20) | −0.005 | $5s^25p$ | $^2P^\circ_{3/2}$ | $5s^26d$ | $^2D_{3/2}$ | 4251.505 | 90241.568 | 1.28+8 | B+ | TW | OH10 |
| 37000 | | 1185.679(10) | 84339.9 | 1185.678(4) | 0.001 | $5s5p^2$ | $^4P_{3/2}$ | $5s5p(^1P^\circ)5d$ | $^2D^\circ_{5/2}$ | 48368.198 | 132708.1 | | | TW | |
| 66000 | bl(Si II) | 1193.289(10) | 83802.0 | 1193.3067(24) | −0.018 | $5s5p^2$ | $^4P_{3/2}$ | $5s5p(^1P^\circ)5d$ | $^2D^\circ_{3/2}$ | 48368.198 | 132168.95 | | | TW | |
| 91000 | | 1219.083(7) | 82028.9 | 1219.08363(22) | −0.001 | $5s^25p$ | $^2P^\circ_{3/2}$ | $5s^27s$ | $^2S_{1/2}$ | 4251.505 | 86280.332 | 3.35+8 | B+ | TW | OH10 |
| 100000 | | 1223.709(7) | 81718.8 | 1223.714(4) | −0.005 | $5s^25p$ | $^2P^\circ_{1/2}$ | $5s5p^2$ | $^2P_{3/2}$ | 0.00 | 81718.43 | 4.08+8 | B+ | TW | OH10 |
| 110000 | | 1242.922(7) | 80455.6 | 1242.926(5) | −0.004 | $5s^25p$ | $^2P^\circ_{1/2}$ | $5s5p^2$ | $^2P_{1/2}$ | 0.00 | 80455.3 | 4.5+8 | B+ | TW | OH10 |
| 31000 | | 1285.659(5) | 77781.1 | 1285.653(3) | 0.006 | $5s5p^2$ | $^4P_{1/2}$ | $5s5p(^3P^\circ)5d$ | $^4P^\circ_{3/2}$ | 46464.301 | 124245.80 | | | TW | |
| 200000 | | 1290.874(7) | 77466.9 | 1290.873(4) | 0.001 | $5s^25p$ | $^2P^\circ_{3/2}$ | $5s5p^2$ | $^2P_{3/2}$ | 4251.505 | 81718.43 | 2.95+9 | B+ | TW | OH10 |
| 37000 | H | 1303.902(10) | 76692.9 | 1303.908(5) | −0.006 | $5s5p^2$ | $^4P_{1/2}$ | $5p3$ | $^4S^\circ_{3/2}$ | 46464.301 | 123156.8 | 9.+8 | D+ | TW | TW |
| 210000 | | 1312.273(7) | 76203.7 | 1312.271(5) | 0.002 | $5s^25p$ | $^2P^\circ_{3/2}$ | $5s5p^2$ | $^2P_{1/2}$ | 4251.505 | 80455.3 | 1.77+9 | B+ | TW | OH10 |

| $I_{obs}$ [a] arb. u. | Char. [b] | $\lambda_{obs}$ [c] Å | $\sigma_{obs}$ cm$^{-1}$ | $\lambda_{Ritz}$ [d] Å | $\delta\lambda_{O-Ritz}$ [e] Å | Classification | | | | $E_{low}$ cm$^{-1}$ | $E_{upp}$ cm$^{-1}$ | $A$ [f] s$^{-1}$ | Acc. [g] | Line Ref. [h] | TP Ref. [h] |
|---|---|---|---|---|---|---|---|---|---|---|---|---|---|---|---|
| 100000 | bl(Sn III) | 1313.090(10) | 76156.2 | 1313.090(10) | | $5s5p^2$ | $^4P_{3/2}$ | $5s5p(^3P°)5d$ | $^4P°_{1/2}$ | 48368.198 | 124524.4 | 2.0+9 | D+ | TW | TW |
| 210000 | | 1316.579(7) | 75954.4 | 1316.579(5) | 0.000 | $5s^25p$ | $^2P°_{1/2}$ | $5s5p^2$ | $^2S_{1/2}$ | 0.00 | 75954.4 | 2.14+9 | B+ | TW | OH10 |
| 86000 | | 1317.906(5) | 75877.9 | 1317.912(3) | −0.006 | $5s5p^2$ | $^4P_{3/2}$ | $5s5p(^3P°)5d$ | $^4P°_{3/2}$ | 48368.198 | 124245.80 | 1.3+9 | D+ | TW | TW |
| 33000 | H | 1327.668(10) | 75320.0 | 1327.667(7) | 0.001 | $5s5p^2$ | $^4P_{3/2}$ | $5s5p(^3P°)5d$ | $^4P°_{5/2}$ | 48368.198 | 123688.3 | | | TW | |
| 67000 | h | 1337.102(10) | 74788.6 | 1337.102(5) | 0.000 | $5s5p^2$ | $^4P_{3/2}$ | $5p3$ | $^4S°_{3/2}$ | 48368.198 | 123156.8 | 1.3+9 | D+ | TW | TW |
| 13000 | H | 1353.843(10) | 73863.8 | 1353.841(5) | 0.002 | $5s5p^2$ | $^2D_{3/2}$ | $5s5p(^1P°)5d$ | $^2D°_{5/2}$ | 58844.194 | 132708.1 | 3.6+8 | D+ | TW | TW |
| 58000 | h | 1358.705(10) | 73599.5 | 1358.692(7) | 0.013 | $5s5p^2$ | $^4P_{1/2}$ | $5s5p(^3P°)5d$ | $^4D°_{3/2}$ | 46464.301 | 120064.5 | 1.9+9 | D+ | TW | TW |
| 200000 | H,bl(Sn III)* | 1360.224(10) | 73517.3 | 1360.256(3) | −0.032 | $5s5p^2$ | $^4P_{5/2}$ | $5s5p(^3P°)5d$ | $^4P°_{3/2}$ | 50730.237 | 124245.80 | 7.5+8 | D+ | TW | TW |
| 120000 | H,bl(Sn III)* | 1360.224(10) | 73517.3 | 1360.230(8) | −0.006 | $5s5p^2$ | $^4P_{1/2}$ | $5s5p(^3P°)5d$ | $^4D°_{1/2}$ | 46464.301 | 119981.3 | 3.3+9 | D+ | TW | TW |
| 93000 | | 1363.797(5) | 73324.7 | 1363.796(3) | 0.001 | $5s5p^2$ | $^2D_{3/2}$ | $5s5p(^1P°)5d$ | $^2D°_{3/2}$ | 58844.194 | 132168.95 | 3.6+9 | D+ | TW | TW |
| 100000 | H | 1365.278(10) | 73245.2 | 1365.288(5) | −0.010 | $5s5p^2$ | $^2D_{5/2}$ | $5s5p(^1P°)5d$ | $^2D°_{5/2}$ | 59463.494 | 132708.1 | 3.4+9 | D+ | TW | TW |
| 100000 | H,bl(Sn III) | 1370.649(10) | 72958.1 | 1370.650(7) | −0.001 | $5s5p^2$ | $^4P_{5/2}$ | $5s5p(^3P°)5d$ | $^4P°_{5/2}$ | 50730.237 | 123688.3 | 2.1+9 | D+ | TW | TW |
| 46000 | | 1375.413(5) | 72705.4 | 1375.413(3) | 0.000 | $5s5p^2$ | $^2D_{5/2}$ | $5s5p(^1P°)5d$ | $^2D°_{3/2}$ | 59463.494 | 132168.95 | 1.8+8 | D+ | TW | TW |
| 68000 | H | 1380.701(10) | 72427.0 | 1380.709(6) | −0.008 | $5s5p^2$ | $^4P_{5/2}$ | $5p3$ | $^4S°_{3/2}$ | 50730.237 | 123156.8 | 2.3+9 | D+ | TW | TW |
| 130000 | s | 1391.097(7) | 71885.7 | 1391.098(4) | −0.001 | $5s5p^2$ | $^4P_{3/2}$ | $5s5p(^3P°)5d$ | $^4D°_{5/2}$ | 48368.198 | 120253.85 | 2.3+9 | D+ | TW | TW |
| 130000 | H | 1393.507(10) | 71761.4 | 1393.507(10) | | $5s5p^2$ | $^4P_{5/2}$ | $5s5p(^3P°)5d$ | $^4D°_{7/2}$ | 50730.237 | 122491.6 | 3.1+9 | D+ | TW | TW |
| 62000 | | 1394.667(19) | 71701.7 | 1394.644(5) | 0.023 | $5s^25p$ | $^2P°_{3/2}$ | $5s5p^2$ | $^2S_{1/2}$ | 4251.505 | 75954.4 | 7.+6 | E | Wu* | OH10 |
| 19000 | H | 1394.761(10) | 71696.9 | 1394.772(7) | −0.011 | $5s5p^2$ | $^4P_{3/2}$ | $5s5p(^3P°)5d$ | $^4D°_{3/2}$ | 48368.198 | 120064.5 | 1.0+9 | D+ | TW | TW |
| 15000 | H | 1396.396(10) | 71612.9 | 1396.393(8) | 0.003 | $5s5p^2$ | $^4P_{3/2}$ | $5s5p(^3P°)5d$ | $^4D°_{1/2}$ | 48368.198 | 119981.3 | 2.4+8 | D+ | TW | TW |
| 360000 | | 1400.448(10) | 71405.7 | 1400.4395(9) | 0.009 | $5s^25p$ | $^2P°_{1/2}$ | $5s^25d$ | $^2D_{3/2}$ | 0.00 | 71406.155 | 2.05+9 | B+ | TW | OH10 |
| 5000 | | 1438.362(10) | 69523.5 | 1438.360(5) | 0.002 | $5s5p^2$ | $^4P_{5/2}$ | $5s5p(^3P°)5d$ | $^4D°_{5/2}$ | 50730.237 | 120253.85 | | | TW | |
| 530000 | | 1474.995(10) | 67796.8 | 1474.9966(3) | −0.002 | $5s^25p$ | $^2P°_{3/2}$ | $5s^25d$ | $^2D_{5/2}$ | 4251.505 | 72048.273 | 1.95+9 | B+ | TW | OH10 |
| 15000 | | 1481.739(5) | 67488.27 | 1481.737(3) | 0.002 | $5s5p^2$ | $^4P_{1/2}$ | $5s5p(^3P°)6s$ | $^2P°_{1/2}$ | 46464.301 | 113952.66 | | | TW | |
| 79000 | | 1489.094(10) | 67154.9 | 1489.1002(4) | −0.006 | $5s^25p$ | $^2P°_{3/2}$ | $5s^25d$ | $^2D_{3/2}$ | 4251.505 | 71406.155 | 1.59+8 | B+ | TW | OH10 |
| 32000 | | 1495.033(5) | 66888.16 | 1495.033(5) | | $5s5p^2$ | $^4P_{3/2}$ | $5s5p(^3P°)5d$ | $^4F°_{5/2}$ | 48368.198 | 115256.35 | 1.1+8 | D+ | TW | TW |
| 33000 | | 1517.957(5) | 65878.02 | 1517.962(3) | −0.005 | $5s5p^2$ | $^4P_{3/2}$ | $5s5p(^3P°)6s$ | $^4P°_{5/2}$ | 48368.198 | 114245.98 | 4.4+8 | D+ | TW | TW |
| 24000 | bl(Sn) | 1522.210(10) | 65694.0 | 1522.228(6) | −0.018 | $5s5p^2$ | $^4P_{5/2}$ | $5s5p(^3P°)5d$ | $^4F°_{7/2}$ | 50730.237 | 116423.4 | 8.3+7 | D+ | TW | TW |
| 15000 | bl(Sn) | 1527.860(10) | 65451.0 | 1527.873(4) | −0.013 | $5s5p^2$ | $^4P_{3/2}$ | $5s5p(^3P°)5d$ | $^2D°_{5/2}$ | 48368.198 | 113818.68 | | | TW | |
| 14000 | | 1543.653(19) | 64781.4 | 1543.631(4) | 0.022 | $5s5p^2$ | $^2D_{5/2}$ | $5s5p(^3P°)5d$ | $^4P°_{3/2}$ | 59463.494 | 124245.80 | | | Wu* | |
| 60000 | | 1554.878(10) | 64313.7 | 1554.882(3) | −0.004 | $5s5p^2$ | $^4P_{1/2}$ | $5s5p(^3P°)6s$ | $^4P°_{3/2}$ | 46464.301 | 110777.85 | 4.0+8 | D+ | TW | TW |
| 13000 | | 1570.056(19) | 63692.0 | 1570.024(7) | 0.032 | $5s5p^2$ | $^2D_{5/2}$ | $5p3$ | $^4S°_{3/2}$ | 59463.494 | 123156.8 | | | Wu* | |
| 33000 | | 1574.417(5) | 63515.57 | 1574.413(4) | 0.004 | $5s5p^2$ | $^4P_{5/2}$ | $5s5p(^3P°)6s$ | $^4P°_{5/2}$ | 50730.237 | 114245.98 | 7.2+8 | D+ | TW | TW |
| 14000 | bl(Sn) | 1585.063(10) | 63089.0 | 1585.076(4) | −0.013 | $5s5p^2$ | $^4P_{5/2}$ | $5s5p(^3P°)5d$ | $^2D°_{5/2}$ | 50730.237 | 113818.68 | 6.2+7 | D+ | TW | TW |

| $I_{obs}$ [a] arb. u. | Char. [b] | $\lambda_{obs}$ [c] Å | $\sigma_{obs}$ cm$^{-1}$ | $\lambda_{Ritz}$ [d] Å | $\delta\lambda_{O-Ritz}$ [e] Å | Classification | | | | $E_{low}$ cm$^{-1}$ | $E_{upp}$ cm$^{-1}$ | $A$ [f] s$^{-1}$ | Acc. [g] | Line Ref. [h] | TP Ref. [h] |
|---|---|---|---|---|---|---|---|---|---|---|---|---|---|---|---|
| 9200 | | 1587.532(19) | 62990.9 | 1587.560(5) | −0.028 | 5s5p$^2$ | $^4P_{1/2}$ | 5s5p($^3$P°)6s | $^4P^\circ_{1/2}$ | 46464.301 | 109454.05 | 1.5+8 | D+ | Wu | TW |
| 15000 | bl(Sn) | 1593.418(10) | 62758.2 | 1593.425(6) | −0.007 | 5s5p$^2$ | $^4P_{1/2}$ | 5s$^2$10p | $^2P^\circ_{1/2}$ | 46464.301 | 109222.18 | | | TW | |
| 16000 | | 1602.313(5) | 62409.78 | 1602.316(3) | −0.003 | 5s5p$^2$ | $^4P_{3/2}$ | 5s5p($^3$P°)6s | $^4P^\circ_{3/2}$ | 48368.198 | 110777.85 | 1.1+8 | D+ | TW | TW |
| 22000 | bl(Sn III) | 1628.409(10) | 61409.6 | 1628.408(6) | 0.001 | 5s5p$^2$ | $^2D_{3/2}$ | 5s5p($^3$P°)5d | $^4D^\circ_{5/2}$ | 58844.194 | 120253.85 | | | TW | |
| 16000 | | 1637.042(5) | 61085.79 | 1637.040(5) | 0.002 | 5s5p$^2$ | $^4P_{3/2}$ | 5s5p($^3$P°)6s | $^4P^\circ_{1/2}$ | 48368.198 | 109454.05 | 6.6+8 | D+ | TW | TW |
| 8100 | | 1643.294(10) | 60853.4 | 1643.278(6) | 0.016 | 5s5p$^2$ | $^4P_{3/2}$ | 5s$^2$10p | $^2P^\circ_{1/2}$ | 48368.198 | 109222.18 | 1.4+8 | D+ | TW | TW |
| 6300 | | 1648.545(10) | 60659.6 | 1648.538(7) | 0.007 | 5s$^2$5d | $^2D_{5/2}$ | 5s5p($^1$P°)5d | $^2D^\circ_{5/2}$ | 72048.273 | 132708.1 | | | TW | |
| 19000 | | 1665.346(5) | 60047.58 | 1665.345(3) | 0.001 | 5s5p$^2$ | $^4P_{5/2}$ | 5s5p($^3$P°)6s | $^4P^\circ_{3/2}$ | 50730.237 | 110777.85 | 3.8+8 | D+ | TW | TW |
| 21000 | | 1699.403(10) | 58844.2 | 1699.4030(13) | 0.000 | 5s$^2$5p | $^2P^\circ_{1/2}$ | 5s5p$^2$ | $^2D_{3/2}$ | 0.00 | 58844.194 | 2.99+7 | B+ | TW | OH10 |
| 1700 | | 1755.621(19) | 56959.9 | 1755.621(8) | 0.000 | 5s5p$^2$ | $^2D_{5/2}$ | 5s5p($^3$P°)5d | $^4F^\circ_{7/2}$ | 59463.494 | 116423.4 | | | Wu* | |
| 15000 | | 1757.891(10) | 56886.3 | 1757.8901(14) | 0.001 | 5s$^2$5p | $^2P^\circ_{1/2}$ | 5s$^2$6s | $^2S_{1/2}$ | 0.00 | 56886.3763 | 3.04+8 | B+ | TW | OH10 |
| 6500 | bl(Sn) | 1778.904(10) | 56214.4 | 1778.899(8) | 0.005 | 5s5p$^2$ | $^2S_{1/2}$ | 5s5p($^1$P°)5d | $^2D^\circ_{3/2}$ | 75954.4 | 132168.95 | | | TW | |
| 1900 | | 1805.002(19) | 55401.6 | 1804.996(5) | 0.006 | 5s5p$^2$ | $^2D_{3/2}$ | 5s5p($^3$P°)6s | $^4P^\circ_{5/2}$ | 58844.194 | 114245.98 | | | Wu | |
| 14000 | | 1811.210(10) | 55211.7 | 1811.2008(5) | 0.009 | 5s$^2$5p | $^2P^\circ_{3/2}$ | 5s5p$^2$ | $^2D_{5/2}$ | 4251.505 | 59463.494 | 6.4+7 | B+ | TW | OH10 |
| 1100 | | 1814.602(5) | 55108.50 | 1814.603(4) | −0.001 | 5s5p$^2$ | $^2D_{3/2}$ | 5s5p($^3$P°)6s | $^2P^\circ_{1/2}$ | 58844.194 | 113952.66 | | | TW | |
| 1500 | | 1819.045(10) | 54973.9 | 1819.026(6) | 0.019 | 5s5p$^2$ | $^2D_{3/2}$ | 5s5p($^3$P°)5d | $^2D^\circ_{5/2}$ | 58844.194 | 113818.68 | | | TW | |
| 11000 | | 1831.727(10) | 54593.3 | 1831.7471(5) | −0.020 | 5s$^2$5p | $^2P^\circ_{3/2}$ | 5s5p$^2$ | $^2D_{3/2}$ | 4251.505 | 58844.194 | 2.2+7 | C+ | TW | OH10 |
| 370 | | 1855.604(19) | 53890.8 | 1855.581(4) | 0.023 | 5s$^2$6s | $^2S_{1/2}$ | 5s5p($^3$P°)6s | $^4P^\circ_{3/2}$ | 56886.3763 | 110777.85 | | | Wu | |
| 570 | | 1886.142(19) | 53018.3 | 1886.117(8) | 0.025 | 5s5p$^2$ | $^4P_{3/2}$ | 5s$^2$8p | $^2P^\circ_{3/2}$ | 48368.198 | 101387.18 | | | Wu | |
| 7200 | | 1899.890(10) | 52634.6 | 1899.8812(5) | 0.009 | 5s$^2$5p | $^2P^\circ_{3/2}$ | 5s$^2$6s | $^2S_{1/2}$ | 4251.505 | 56886.3763 | 5.6+8 | B+ | TW | OH10 |
| 170 | | 2108.475(19) | 47412.6 | 2108.493(12) | −0.018 | 5s5p$^2$ | $^2D_{3/2}$ | 5s$^2$9p | $^2P^\circ_{1/2}$ | 58844.194 | 106256.4 | | | Wu | |
| 250 | | 2131.208(19) | 46906.9 | 2131.219(18) | −0.011 | 5s5p$^2$ | $^2D_{5/2}$ | 5s$^2$9p | $^2P^\circ_{3/2}$ | 59463.494 | 106370.2 | | | Wu | |
| 720 | bl(Sn III) | 2148.61(8) | 46527.1 | 2148.590(16) | 0.02 | 5s5p$^2$ | $^2D_{3/2}$ | 5s$^2$6f | $^2F^\circ_{5/2}$ | 58844.194 | 105371.7 | | | MS | |
| 520 | | 2150.8442(9) | 46478.749 | 2150.8450(7) | −0.0008 | 5s$^2$5p | $^2P^\circ_{3/2}$ | 5s5p$^2$ | $^4P_{5/2}$ | 4251.505 | 50730.237 | 4.0+5 | D+ | Brill | OH10 |
| 530 | | 2151.5135(20) | 46464.29 | 2151.5131(19) | 0.0004 | 5s$^2$5p | $^2P^\circ_{1/2}$ | 5s5p$^2$ | $^4P_{1/2}$ | 0.00 | 46464.301 | 2.1+6 | C+ | Brill | OH10 |
| 160 | | 2200.075(19) | 45438.8 | 2200.0340(11) | 0.041 | 5s5p$^2$ | $^4P_{1/2}$ | 5s$^2$7p | $^2P^\circ_{1/2}$ | 46464.301 | 91903.958 | | | Wu | |
| | m(Sn I) | | | 2246.454(11) | | 5s$^2$6s | $^2S_{1/2}$ | 5s$^2$8p | $^2P^\circ_{3/2}$ | 56886.3763 | 101387.18 | | | MS | |
| 67 | | 2252.845(19) | 44374.6 | 2252.817(13) | 0.028 | 5s$^2$5d | $^2D_{5/2}$ | 5s5p($^3$P°)5d | $^4F^\circ_{7/2}$ | 72048.273 | 116423.4 | | | Wu* | |
| 35 | | 2255.726(19) | 44317.9 | 2255.730(19) | −0.004 | 5s$^2$6s | $^2S_{1/2}$ | 5s$^2$8p | $^2P^\circ_{1/2}$ | 56886.3763 | 101204.2 | | | Wu | |
| 260 | | 2266.0156(10) | 44116.677 | 2266.0148(7) | 0.0008 | 5s$^2$5p | $^2P^\circ_{3/2}$ | 5s5p$^2$ | $^4P_{3/2}$ | 4251.505 | 48368.198 | 4.3+5 | D+ | Brill | OH10 |
| 86 | | 2296.293(19) | 43535.0 | 2296.2548(9) | 0.038 | 5s5p$^2$ | $^4P_{3/2}$ | 5s$^2$7p | $^2P^\circ_{1/2}$ | 48368.198 | 91903.958 | | | Wu | |
| 45 | | 2333.43(14) | 42842 | 2333.561(8) | −0.13 | 5s$^2$5d | $^2D_{3/2}$ | 5s5p($^3$P°)6s | $^4P^\circ_{5/2}$ | 71406.155 | 114245.98 | | | Wu | |
| 120 | | 2349.825(19) | 42543.3 | 2349.844(12) | −0.019 | 5s5p$^2$ | $^2D_{3/2}$ | 5s$^2$8p | $^2P^\circ_{3/2}$ | 58844.194 | 101387.18 | | | Wu | |

| $I_{obs}$ [a] arb. u. | Char. [b] | $\lambda_{obs}$ [c] Å | $\sigma_{obs}$ $cm^{-1}$ | $\lambda_{Ritz}$ [d] Å | $\delta\lambda_{O-Ritz}$ [e] Å | Classification | | | | $E_{low}$ $cm^{-1}$ | $E_{upp}$ $cm^{-1}$ | $A$ [f] $s^{-1}$ | Acc. [g] | Line Ref. [h] | TP Ref. [h] |
|---|---|---|---|---|---|---|---|---|---|---|---|---|---|---|---|
| 130 | | 2350.698(19) | 42527.5 | 2350.707(14) | −0.009 | $5s5p^2$ | $^2P_{3/2}$ | $5s5p(^3P^\circ)5d$ | $^4P^\circ_{3/2}$ | 81718.43 | 124245.80 | | | Wu | |
| | m | | | 2357.073(10) | | $5s^25d$ | $^2D_{3/2}$ | $5s5p(^3P^\circ)5d$ | $^2D^\circ_{5/2}$ | 71406.155 | 113818.68 | | | Wu | |
| | m | | | 2359.996(21) | | $5s5p^2$ | $^2D_{3/2}$ | $5s^28p$ | $^2P^\circ_{1/2}$ | 58844.194 | 101204.2 | | | Wu | |
| 3700 | | 2360.28(10) | 42355 | 2360.208(14) | 0.07x | $5s^26d$ | $^2D_{5/2}$ | $5s5p(^1P^\circ)5d$ | $^2D^\circ_{5/2}$ | 90351.908 | 132708.1 | | | Wu* | |
| 310 | | 2368.2265(6) | 42212.795 | 2368.2265(6) | 0.0000 | $5s^25p$ | $^2P^\circ_{3/2}$ | $5s5p^2$ | $^4P_{1/2}$ | 4251.505 | 46464.301 | 5.6+5 | C+ | Brill | OH10 |
| 48 | | 2369.15(10) | 42196.4 | 2369.073(8) | 0.08 | $5s^25d$ | $^2D_{5/2}$ | $5s5p(^3P^\circ)6s$ | $^4P^\circ_{5/2}$ | 72048.273 | 114245.98 | | | Wu | |
| 220 | | 2384.565(19) | 41923.6 | 2384.559(12) | 0.006 | $5s5p^2$ | $^2D_{5/2}$ | $5s^28p$ | $^2P^\circ_{3/2}$ | 59463.494 | 101387.18 | | | Wu | |
| 53 | | 2393.309(19) | 41770.4 | 2393.310(10) | −0.001 | $5s^25d$ | $^2D_{5/2}$ | $5s5p(^3P^\circ)5d$ | $^2D^\circ_{5/2}$ | 72048.273 | 113818.68 | | | Wu* | |
| 85 | | 2406.712(19) | 41537.8 | 2406.7088(6) | 0.003 | $5s5p^2$ | $^4P_{5/2}$ | $5s^27p$ | $^2P^\circ_{3/2}$ | 50730.237 | 92268.119 | 3.4+5 | D+ | Wu | OH10 |
| 160 | bl(Sn I) | 2433.48(3) | 41080.9 | 2433.49(3) | −0.01 | $5s^26p$ | $^2P^\circ_{1/2}$ | $5s^211d$ | $^2D_{3/2}$ | 71493.287 | 112574.1 | | | Wu | |
| 5700 | : | 2442.7 | 40926 | 2442.7019(6) | | $5s5p^2$ | $^4P_{3/2}$ | $5s^24f$ | $^2F^\circ_{5/2}$ | 48368.198 | 89294.068 | 2.4+5 | D+ | AM | OH10 |
| 3500 | | 2448.9079(7) | 40822.163 | 2448.9089(4) | −0.0010 | $5s5p^2$ | $^2D_{3/2}$ | $5s^25f$ | $^2F^\circ_{5/2}$ | 58844.194 | 99666.340 | 6.4+7 | D+ | Brill | TW |
| 2300 | : | 2486.6 | 40203 | 2486.6356(5) | | $5s5p^2$ | $^2D_{5/2}$ | $5s^25f$ | $^2F^\circ_{5/2}$ | 59463.494 | 99666.340 | 1.0+7 | D | AM | AM |
| 4100 | | 2486.9666(8) | 40197.495 | 2486.9665(4) | 0.0001 | $5s5p^2$ | $^2D_{5/2}$ | $5s^25f$ | $^2F^\circ_{7/2}$ | 59463.494 | 99660.991 | 6.8+7 | D+ | Brill | TW |
| 1700 | | 2522.69(9) | 39628.3 | 2522.63(8) | 0.06 | $5s^26p$ | $^2P^\circ_{1/2}$ | $5s^210d$ | $^2D_{3/2}$ | 71493.287 | 111122.6 | | | MS | |
| 1900 | | 2538.95(10) | 39374.5 | 2539.133(8) | −0.18x | $5s^25d$ | $^2D_{3/2}$ | $5s5p(^3P^\circ)6s$ | $^4P^\circ_{3/2}$ | 71406.155 | 110777.85 | | | Wu | |
| 190 | | 2579.15(23) | 38761 | 2578.82(7) | 0.33 | $5s^26p$ | $^2P^\circ_{3/2}$ | $5s^210d$ | $^2D_{5/2}$ | 72377.4616 | 111143.3 | 1.0+7 | D+ | MS | TW |
| 550 | : | 2592.3 | 38564 | 2592.3281(5) | | $5s5p^2$ | $^4P_{5/2}$ | $5s^24f$ | $^2F^\circ_{5/2}$ | 50730.237 | 89294.068 | 1.9+5 | D+ | AM | OH10 |
| 1200 | | 2592.7198(17) | 38558.01 | 2592.7181(5) | 0.0017 | $5s5p^2$ | $^4P_{5/2}$ | $5s^24f$ | $^2F^\circ_{7/2}$ | 50730.237 | 89288.268 | 2.9+6 | C+ | Brill | OH10 |
| 200 | | 2608.74(24) | 38321 | 2608.74(24) | | $5s^26p$ | $^2P^\circ_{3/2}$ | $5s^211s$ | $^2S_{1/2}$ | 72377.4616 | 110699 | | | MS | |
| 880 | bl(Sn III) | 2643.56(3) | 37816.5 | 2643.594(16) | −0.03 | $5s^25d$ | $^2D_{3/2}$ | $5s^210p$ | $^2P^\circ_{1/2}$ | 71406.155 | 109222.18 | | | Wu | |
| 860 | | 2664.99(10) | 37512.4 | 2664.96(3) | 0.03 | $5s^26p$ | $^2P^\circ_{1/2}$ | $5s^29d$ | $^2D_{3/2}$ | 71493.287 | 109006.2 | 1.4+7 | D+ | MS | TW |
| 220 | | 2711.86(3) | 36864.2 | 2711.85(3) | 0.01 | $5s^26p$ | $^2P^\circ_{1/2}$ | $5s^210s$ | $^2S_{1/2}$ | 71493.287 | 108357.6 | | | Wu | |
| 400 | | 2727.76(3) | 36649.3 | 2727.838(11) | −0.08 | $5s^26p$ | $^2P^\circ_{3/2}$ | $5s^29d$ | $^2D_{5/2}$ | 72377.4616 | 109025.68 | 1.6+7 | D+ | Wu | TW |
| 180 | | 2778.4(3) | 35982 | 2778.49(3) | −0.1 | $5s^26p$ | $^2P^\circ_{3/2}$ | $5s^210s$ | $^2S_{1/2}$ | 72377.4616 | 108357.6 | | | MS | |
| 240 | | 2825.51(3) | 35381.4 | 2825.4849(7) | 0.03 | $5s^26s$ | $^2S_{1/2}$ | $5s^27p$ | $^2P^\circ_{3/2}$ | 56886.3763 | 92268.119 | 1.11+6 | C+ | Wu | OH10 |
| 210 | | 2868.61(3) | 34849.9 | 2868.578(23) | 0.03 | $5s^25d$ | $^2D_{3/2}$ | $5s^29p$ | $^2P^\circ_{1/2}$ | 71406.155 | 106256.4 | | | Wu | |
| 220 | | 2912.82(10) | 34320.9 | 2912.74(3) | 0.08 | $5s^25d$ | $^2D_{5/2}$ | $5s^29p$ | $^2P^\circ_{3/2}$ | 72048.273 | 106370.2 | | | MS | |
| 610 | | 2919.86(3) | 34238.2 | 2919.884(25) | −0.02 | $5s^26p$ | $^2P^\circ_{1/2}$ | $5s^28d$ | $^2D_{3/2}$ | 71493.287 | 105731.2 | 2.4+7 | D+ | Wu | TW |
| 190 | | 2943.30(3) | 33965.5 | 2943.30(3) | 0.00 | $5s^25d$ | $^2D_{3/2}$ | $5s^26f$ | $^2F^\circ_{5/2}$ | 71406.155 | 105371.7 | | | Wu | |
| 400 | | 2949.54(3) | 33893.7 | 2949.522(14) | 0.02 | $5s^26d$ | $^2D_{5/2}$ | $5s5p(^3P^\circ)5d$ | $^4P^\circ_{3/2}$ | 90351.908 | 124245.80 | | | Wu | |
| 220 | | 2990.99(3) | 33424.0 | 2990.9965(8) | −0.01 | $5s5p^2$ | $^2D_{3/2}$ | $5s^27p$ | $^2P^\circ_{3/2}$ | 58844.194 | 92268.119 | 9.1+5 | C+ | Wu | OH10 |
| 790 | | 2994.46(3) | 33385.3 | 2994.434(20) | 0.03 | $5s^26p$ | $^2P^\circ_{3/2}$ | $5s^28d$ | $^2D_{5/2}$ | 72377.4616 | 105763.02 | 2.7+7 | D+ | Wu | TW |

| $I_{obs}$ [a] arb. u. | Char. [b] | $\lambda_{obs}$ [c] Å | $\sigma_{obs}$ cm$^{-1}$ | $\lambda_{Ritz}$ [d] Å | $\delta\lambda_{O-Ritz}$ [e] Å | Classification | | | | $E_{low}$ cm$^{-1}$ | $E_{upp}$ cm$^{-1}$ | $A$ [f] s$^{-1}$ | Acc. [g] | Line Ref. [h] | TP Ref. [h] |
|---|---|---|---|---|---|---|---|---|---|---|---|---|---|---|---|
| 110 | | 2997.1(3) | 33355 | 2997.29(3) | −0.2 | 5s$^2$6p | $^2$P°$_{3/2}$ | 5s$^2$8d | $^2$D$_{3/2}$ | 72377.4616 | 105731.2 | | | MS | |
| 260 | | 3012.41(5) | 33186.3 | 3012.519(9) | −0.11 | 5s$^2$6p | $^2$P°$_{1/2}$ | 5s$^2$9s | $^2$S$_{1/2}$ | 71493.287 | 104678.43 | | | Wu | |
| 450 | | 3023.92(3) | 33060.1 | 3023.9444(14) | −0.02 | 5s5p$^2$ | $^2$D$_{3/2}$ | 5s$^2$7p | $^2$P°$_{1/2}$ | 58844.194 | 91903.958 | 7.8+6 | C+ | Wu | OH10 |
| 680 | | 3047.44(3) | 32804.9 | 3047.4642(9) | −0.02 | 5s5p$^2$ | $^2$D$_{5/2}$ | 5s$^2$7p | $^2$P°$_{3/2}$ | 59463.494 | 92268.119 | 6.8+6 | B+ | Wu | OH10 |
| 930 | | 3094.68(11) | 32304.1 | 3094.984(9) | −0.30 | 5s$^2$6p | $^2$P°$_{3/2}$ | 5s$^2$9s | $^2$S$_{1/2}$ | 72377.4616 | 104678.43 | | | MS | |
| 480 | | 3101.25(16) | 32235.7 | 3101.39(3) | −0.14x | 5s5p$^2$ | $^2$P$_{3/2}$ | 5s5p($^3$P°)6s | $^2$P°$_{1/2}$ | 81718.43 | 113952.66 | | | Wu* | |
| 15000 | | 3283.1399(9) | 30449.874 | 3283.1399(7) | 0.0000 | 5s5p$^2$ | $^2$D$_{3/2}$ | 5s$^2$4f | $^2$F°$_{5/2}$ | 58844.194 | 89294.068 | 1.70+8 | B+ | Brill | OH10 |
| 13000 | : | 3351.3 | 29830.6 | 3351.3021(8) | | 5s5p$^2$ | $^2$D$_{5/2}$ | 5s$^2$4f | $^2$F°$_{5/2}$ | 59463.494 | 89294.068 | 1.21+7 | B+ | AM | OH10 |
| 13000 | | 3351.9523(12) | 29824.788 | 3351.9538(8) | −0.0015 | 5s5p$^2$ | $^2$D$_{5/2}$ | 5s$^2$4f | $^2$F°$_{7/2}$ | 59463.494 | 89288.268 | 1.82+8 | B+ | Brill | OH10 |
| 87 | | 3355.5(3) | 29793 | 3354.96(4) | 0.5 | 5s$^2$5d | $^2$D$_{3/2}$ | 5s$^2$8p | $^2$P°$_{1/2}$ | 71406.155 | 101204.2 | | | MS | |
| 350 | | 3407.41(12) | 29339.4 | 3407.466(25) | −0.06 | 5s$^2$5d | $^2$D$_{5/2}$ | 5s$^2$8p | $^2$P°$_{3/2}$ | 72048.273 | 101387.18 | | | MS | |
| 1700 | | 3472.333(3) | 28790.837 | 3472.3329(12) | 0.000 | 5s$^2$6p | $^2$P°$_{1/2}$ | 5s$^2$7d | $^2$D$_{3/2}$ | 71493.287 | 100284.125 | 4.5+7 | D+ | Brill | TW |
| 660 | | 3537.47(13) | 28260.7 | 3537.5363(12) | −0.07 | 5s$^2$5d | $^2$D$_{3/2}$ | 5s$^2$5f | $^2$F°$_{5/2}$ | 71406.155 | 99666.340 | 3.6+6 | D | MS | AM |
| 2100 | | 3575.3255(12) | 27961.499 | 3575.3255(12) | 0.0000 | 5s$^2$6p | $^2$P°$_{3/2}$ | 5s$^2$7d | $^2$D$_{5/2}$ | 72377.4616 | 100338.960 | 5.0+7 | D+ | Brill | TW |
| 440 | | 3582.3511(14) | 27906.663 | 3582.3510(13) | 0.0001 | 5s$^2$6p | $^2$P°$_{3/2}$ | 5s$^2$7d | $^2$D$_{3/2}$ | 72377.4616 | 100284.125 | 8.3+6 | D+ | Brill | TW |
| 110 | | 3612.68(22) | 27672.4 | 3612.688(16) | −0.01 | 5s$^2$7s | $^2$S$_{1/2}$ | 5s5p($^3$P°)6s | $^2$P°$_{1/2}$ | 86280.332 | 113952.66 | | | Wu* | |
| 240 | | 3619.96(13) | 27616.7 | 3619.7860(12) | 0.17 | 5s$^2$5d | $^2$D$_{5/2}$ | 5s$^2$5f | $^2$F°$_{5/2}$ | 72048.273 | 99666.340 | 6.2+6 | D | MS | AM |
| 530 | | 3620.4854(15) | 27612.732 | 3620.4872(10) | −0.0018 | 5s$^2$5d | $^2$D$_{5/2}$ | 5s$^2$5f | $^2$F°$_{7/2}$ | 72048.273 | 99660.991 | 2.0+6 | E | Brill | M79 |
| 500 | | 3715.1524(11) | 26909.141 | 3715.1529(9) | −0.0005 | 5s$^2$6p | $^2$P°$_{1/2}$ | 5s$^2$8s | $^2$S$_{1/2}$ | 71493.287 | 98402.425 | 1.6+7 | D+ | Brill | TW |
| 440 | | 3841.3756(14) | 26024.959 | 3841.3750(9) | 0.0006 | 5s$^2$6p | $^2$P°$_{3/2}$ | 5s$^2$8s | $^2$S$_{1/2}$ | 72377.4616 | 98402.425 | 2.9+7 | D+ | Brill | TW |
| 190 | * | 3984.6(4) | 25089.8 | 3984.6(4) | | 5s$^2$4f | $^2$F°$_{7/2}$ | 5s$^2$11g | $^2$G$_{9/2}$ | 89288.268 | 114378.1 | 2.4+6 | D+ | MS | TW |
| 190 | * | 3984.6(4) | 25089.8 | 3984.6(4) | | 5s$^2$4f | $^2$F°$_{7/2}$ | 5s$^2$11g | $^2$G$_{7/2}$ | 89288.268 | 114378.1 | 8.+4 | E | MS | TW |
| 17 | | 3994.3(4) | 25028.7 | 3994.238(3) | 0.1 | 5s5p$^2$ | $^4$P$_{1/2}$ | 5s$^2$6p | $^2$P°$_{1/2}$ | 46464.301 | 71493.287 | | | MS | |
| 180 | * | 4110.3(4) | 24322.6 | 4110.3(3) | 0.0 | 5s$^2$4f | $^2$F°$_{7/2}$ | 5s$^2$10g | $^2$G$_{9/2}$ | 89288.268 | 113610.5 | 3.4+6 | D+ | MS | TW |
| 180 | * | 4110.3(4) | 24322.6 | 4110.3(3) | 0.0 | 5s$^2$4f | $^2$F°$_{7/2}$ | 5s$^2$10g | $^2$G$_{7/2}$ | 89288.268 | 113610.5 | 1.2+5 | E | MS | TW |
| 180 | | 4111.3(4) | 24316.2 | 4111.3(3) | 0.0 | 5s$^2$4f | $^2$F°$_{5/2}$ | 5s$^2$10g | $^2$G$_{7/2}$ | 89294.068 | 113610.5 | 3.3+6 | D+ | MS | TW |
| 28 | | 4164.8(3) | 24004.0 | 4164.727(25) | 0.1 | 5s$^2$6d | $^2$D$_{3/2}$ | 5s5p($^3$P°)6s | $^4$P°$_{5/2}$ | 90241.568 | 114245.98 | | | Wu | |
| 31 | | 4172.2(3) | 23961.4 | 4172.15(3) | 0.1 | 5s$^2$7d | $^2$D$_{3/2}$ | 5s5p($^3$P°)5d | $^4$P°$_{3/2}$ | 100284.125 | 124245.80 | | | Wu | |
| 36 | bl(Sn IV) | 4216.2(6) | 23711 | 4216.248(22) | 0.0 | 5s$^2$6d | $^2$D$_{3/2}$ | 5s5p($^3$P°)6s | $^2$P°$_{1/2}$ | 90241.568 | 113952.66 | | | Wu* | |
| 170 | * | 4293.3(4) | 23285.6 | 4293.27(14) | 0.0 | 5s$^2$4f | $^2$F°$_{7/2}$ | 5s$^2$9g | $^2$G$_{9/2}$ | 89288.268 | 112574.0 | 5.1+6 | D+ | MS | TW |
| 170 | * | 4293.3(4) | 23285.6 | 4293.27(14) | 0.0 | 5s$^2$4f | $^2$F°$_{7/2}$ | 5s$^2$9g | $^2$G$_{7/2}$ | 89288.268 | 112574.0 | 1.8+5 | E | MS | TW |
| 340 | | 4294.33(15) | 23280.0 | 4294.34(14) | −0.01 | 5s$^2$4f | $^2$F°$_{5/2}$ | 5s$^2$9g | $^2$G$_{7/2}$ | 89294.068 | 112574.0 | 4.9+6 | D+ | MS | TW |
| 6 | | 4323.0925(13) | 23125.086 | 4323.0920(13) | 0.0005 | 5s5p$^2$ | $^4$P$_{3/2}$ | 5s$^2$6p | $^2$P°$_{1/2}$ | 48368.198 | 71493.287 | 7.0+4 | D+ | Brill | OH10 |

| $I_{obs}$ [a] arb. u. | Char. [b] | $\lambda_{obs}$ [c] Å | $\sigma_{obs}$ cm$^{-1}$ | $\lambda_{Ritz}$ [d] Å | $\delta\lambda_{O-Ritz}$ [e] Å | Classification | | | | $E_{low}$ cm$^{-1}$ | $E_{upp}$ cm$^{-1}$ | $A$ [f] s$^{-1}$ | Acc. [g] | Line Ref. [h] | TP Ref. [h] |
|---|---|---|---|---|---|---|---|---|---|---|---|---|---|---|---|
| 120 | | 4573.7(4) | 21858.0 | 4574.32(23) | −0.6 | 5s$^2$4f | $^2$F°$_{7/2}$ | 5s$^2$10d | $^2$D$_{5/2}$ | 89288.268 | 111143.3 | | | Wu | |
| 91 | | 4575.0(4) | 21851.8 | 4575.54(23) | −0.5 | 5s$^2$4f | $^2$F°$_{5/2}$ | 5s$^2$10d | $^2$D$_{5/2}$ | 89294.068 | 111143.3 | | | Wu | |
| | m | | | 4579.9(3) | | 5s$^2$4f | $^2$F°$_{5/2}$ | 5s$^2$10d | $^2$D$_{3/2}$ | 89294.068 | 111122.6 | | | Wu | |
| 150 | * | 4579.06(13) | 21832.4 | 4579.04(9) | 0.02 | 5s$^2$4f | $^2$F°$_{7/2}$ | 5s$^2$8g | $^2$G$_{9/2}$ | 89288.268 | 111120.8 | 8.0+6 | D+ | MS | TW |
| 150 | * | 4579.06(13) | 21832.4 | 4579.04(9) | 0.02 | 5s$^2$4f | $^2$F°$_{7/2}$ | 5s$^2$8g | $^2$G$_{7/2}$ | 89288.268 | 111120.8 | 2.9+5 | E | MS | TW |
| 140 | | 4580.22(13) | 21826.9 | 4580.25(9) | −0.03 | 5s$^2$4f | $^2$F°$_{5/2}$ | 5s$^2$8g | $^2$G$_{7/2}$ | 89294.068 | 111120.8 | 7.7+6 | D+ | MS | TW |
| 90 | | 4618.2359(10) | 21647.226 | 4618.2363(10) | −0.0004 | 5s5p$^2$ | $^4$P$_{5/2}$ | 5s$^2$6p | $^2$P°$_{3/2}$ | 50730.237 | 72377.4616 | 6.4+5 | C+ | Brill | OH10 |
| 48 | | 4776.1(4) | 20931.7 | 4776.07(8) | 0.0 | 5s5p$^2$ | $^2$P$_{1/2}$ | 5s$^2$8p | $^2$P°$_{3/2}$ | 80455.3 | 101387.18 | | | Wu | |
| 62 | h | 4792.0732(19) | 20861.963 | 4792.0730(15) | 0.0002 | 5s$^2$5d | $^2$D$_{3/2}$ | 5s$^2$7p | $^2$P°$_{3/2}$ | 71406.155 | 92268.119 | 4.0+5 | C+ | Brill | OH10 |
| 100 | | 4877.209(3) | 20497.805 | 4877.209(3) | 0.000 | 5s$^2$5d | $^2$D$_{3/2}$ | 5s$^2$7p | $^2$P°$_{1/2}$ | 71406.155 | 91903.958 | 5.6+6 | B+ | Brill | OH10 |
| 66 | | 4895.1(4) | 20422.9 | 4894.37(3) | 0.7 | 5s$^2$6d | $^2$D$_{5/2}$ | 5s5p($^3$P°)6s | $^4$P°$_{3/2}$ | 90351.908 | 110777.85 | | | Wu | |
| 83 | | 4917.1(4) | 20331.5 | 4917.8(3) | −0.7 | 5s$^2$7p | $^2$P°$_{3/2}$ | 5s$^2$11d | $^2$D$_{5/2}$ | 92268.119 | 112596.9 | 3.4+6 | D+ | Wu | TW |
| 150 | | 4944.2562(20) | 20219.846 | 4944.2561(16) | 0.0001 | 5s$^2$5d | $^2$D$_{5/2}$ | 5s$^2$7p | $^2$P°$_{3/2}$ | 72048.273 | 92268.119 | 4.9+6 | B+ | Brill | OH10 |
| 360 | * | 5071.09(15) | 19714.1 | 5071.12(11) | −0.03 | 5s$^2$4f | $^2$F°$_{7/2}$ | 5s$^2$7g | $^2$G$_{9/2}$ | 89288.268 | 109002.3 | 1.4+7 | D+ | MS | TW |
| 360 | * | 5071.09(15) | 19714.1 | 5071.12(11) | −0.03 | 5s$^2$4f | $^2$F°$_{7/2}$ | 5s$^2$7g | $^2$G$_{7/2}$ | 89288.268 | 109002.3 | 5.+5 | E | MS | TW |
| | m | | | 5071.60(10) | | 5s$^2$4f | $^2$F°$_{5/2}$ | 5s$^2$9d | $^2$D$_{3/2}$ | 89294.068 | 109006.2 | | | Wu | |
| 360 | | 5072.62(15) | 19708.2 | 5072.61(11) | 0.01 | 5s$^2$4f | $^2$F°$_{5/2}$ | 5s$^2$7g | $^2$G$_{7/2}$ | 89294.068 | 109002.3 | 1.3+7 | D+ | MS | TW |
| 1600 | | 5332.3391(16) | 18748.281 | 5332.3391(11) | 0.0000 | 5s$^2$6p | $^2$P°$_{1/2}$ | 5s$^2$6d | $^2$D$_{3/2}$ | 71493.287 | 90241.568 | 9.9+7 | B+ | Brill | OH10 |
| 2700 | | 5561.9101(16) | 17974.443 | 5561.9091(15) | 0.0010 | 5s$^2$6p | $^2$P°$_{3/2}$ | 5s$^2$6d | $^2$D$_{5/2}$ | 72377.4616 | 90351.908 | 1.13+8 | B+ | Brill | OH10 |
| 2600 | | 5588.8152(18) | 17887.913 | 5588.8153(16) | −0.0001 | 5s$^2$5d | $^2$D$_{3/2}$ | 5s$^2$4f | $^2$F°$_{5/2}$ | 71406.155 | 89294.068 | 7.8+7 | B+ | Brill | OH10 |
| 530 | h | 5596.2644(15) | 17864.103 | 5596.2634(12) | 0.0010 | 5s$^2$6p | $^2$P°$_{3/2}$ | 5s$^2$6d | $^2$D$_{3/2}$ | 72377.4616 | 90241.568 | 1.91+7 | B+ | Brill | OH10 |
| 490 | | 5796.9078(15) | 17245.794 | 5796.9075(13) | 0.0003 | 5s$^2$5d | $^2$D$_{5/2}$ | 5s$^2$4f | $^2$F°$_{5/2}$ | 72048.273 | 89294.068 | 5.1+6 | B+ | Brill | OH10 |
| 2700 | | 5798.860(3) | 17239.988 | 5798.8578(18) | 0.002 | 5s$^2$5d | $^2$D$_{5/2}$ | 5s$^2$4f | $^2$F°$_{7/2}$ | 72048.273 | 89288.268 | 7.7+7 | B+ | Brill | OH10 |
| 470 | H | 5965.84(6) | 16757.46 | 5965.80(5) | 0.04 | 5s$^2$7p | $^2$P°$_{3/2}$ | 5s$^2$9d | $^2$D$_{5/2}$ | 92268.119 | 109025.68 | 7.6+6 | D+ | Brill | TW |
| 1500 | * | 6077.6331(19) | 16449.220 | 6077.6304(16) | 0.0027 | 5s$^2$4f | $^2$F°$_{7/2}$ | 5s$^2$6g | $^2$G$_{9/2}$ | 89288.268 | 105737.495 | 2.7+7 | D+ | Brill | TW |
| 1500 | * | 6077.6331(19) | 16449.220 | 6077.6304(16) | 0.0027 | 5s$^2$4f | $^2$F°$_{7/2}$ | 5s$^2$6g | $^2$G$_{7/2}$ | 89288.268 | 105737.495 | 1.0+6 | D+ | Brill | TW |
| 1400 | | 6079.7696(24) | 16443.439 | 6079.7742(18) | −0.0046 | 5s$^2$4f | $^2$F°$_{5/2}$ | 5s$^2$6g | $^2$G$_{7/2}$ | 89294.068 | 105737.495 | 2.6+7 | D+ | Brill | TW |
| 380 | | 6242.1(7) | 16015.8 | 6241.14(15) | 1.0 | 5s$^2$6d | $^2$D$_{5/2}$ | 5s$^2$9p | $^2$P°$_{3/2}$ | 90351.908 | 106370.2 | | | Wu | |
| 760 | | 6428.4(7) | 15551.7 | 6428.99(5) | −0.6 | 5s$^2$8s | $^2$S$_{1/2}$ | 5s5p($^3$P°)6s | $^2$P°$_{1/2}$ | 98402.425 | 113952.66 | | | Wu* | |
| 2500 | | 6453.5422(12) | 15491.085 | 6453.5421(11) | 0.0001 | 5s$^2$6s | $^2$S$_{1/2}$ | 5s$^2$6p | $^2$P°$_{3/2}$ | 56886.3763 | 72377.4616 | 7.0+7 | B+ | Brill | OH10 |
| 830 | | 6569.7(7) | 15217.2 | 6568.44(9) | 1.3 | 5s$^2$9d | $^2$D$_{5/2}$ | 5s5p($^3$P°)5d | $^4$P°$_{3/2}$ | 109025.68 | 124245.80 | | | Wu | |
| 1000 | | 6661.1(8) | 15008.4 | 6661.1(8) | | 5s$^2$6d | $^2$D$_{5/2}$ | 5s$^2$6f | $^2$F°$_{7/2}$ | 90351.908 | 105360.3 | 1.6+7 | D+ | Wu | TW |
| 840 | | 6760.812(3) | 14787.041 | 6760.8103(22) | 0.002 | 5s$^2$6p | $^2$P°$_{1/2}$ | 5s$^2$7s | $^2$S$_{1/2}$ | 71493.287 | 86280.332 | 3.82+7 | B+ | Brill | OH10 |

| $I_{obs}$ [a] arb. u. | Char. [b] | $\lambda_{obs}$ [c] Å | $\sigma_{obs}$ cm$^{-1}$ | $\lambda_{Ritz}$ [d] Å | $\delta\lambda_{O-Ritz}$ [e] Å | Classification | | | | $E_{low}$ cm$^{-1}$ | $E_{upp}$ cm$^{-1}$ | $A$ [f] s$^{-1}$ | Acc. [g] | Line Ref. [h] | TP Ref. [h] |
|---|---|---|---|---|---|---|---|---|---|---|---|---|---|---|---|
| 1300 | | 6844.1859(20) | 14606.911 | 6844.1860(15) | −0.0001 | $5s^26s$ | $^2S_{1/2}$ | $5s^26p$ | $^2P^\circ_{1/2}$ | 56886.3763 | 71493.287 | 6.0+7 | B+ | Brill | OH10 |
| 1100 | | 7190.776(3) | 13902.873 | 7190.7774(24) | −0.001 | $5s^26p$ | $^2P^\circ_{3/2}$ | $5s^27s$ | $^2S_{1/2}$ | 72377.4616 | 86280.332 | 7.2+7 | B+ | Brill | OH10 |
| 670 | | 7230.1(9) | 13827.3 | 7230.11(15) | 0.0 | $5s^27p$ | $^2P^\circ_{1/2}$ | $5s^28d$ | $^2D_{3/2}$ | 91903.958 | 105731.2 | 1.2+7 | D+ | Wu | TW |
| 500 | | 7314.5(9) | 13667.7 | 7314.06(7) | 0.4 | $5s^27d$ | $^2D_{3/2}$ | $5s5p(^3P^\circ)6s$ | $^2P^\circ_{1/2}$ | 100284.125 | 113952.66 | | | Wu* | |
| 480 | | 7387.1651(24) | 13533.265 | 7387.1636(19) | 0.0015 | $5s5p^2$ | $^2D_{3/2}$ | $5s^26p$ | $^2P^\circ_{3/2}$ | 58844.194 | 72377.4616 | 2.27+6 | B+ | Brill | OH10 |
| 380 | | 7408.22(21) | 13494.8 | 7408.17(12) | 0.05 | $5s^27p$ | $^2P^\circ_{3/2}$ | $5s^28d$ | $^2D_{5/2}$ | 92268.119 | 105763.02 | 1.3+7 | D+ | MS | TW |
| 450 | | 7729.6(10) | 12933.7 | 7728.3(7) | 1.3 | $5s^25f$ | $^2F^\circ_{7/2}$ | $5s^211d$ | $^2D_{5/2}$ | 99660.991 | 112596.9 | | | Wu | |
| 500 | | 7741.425(3) | 12913.965 | 7741.423(3) | 0.002 | $5s5p^2$ | $^2D_{5/2}$ | $5s^26p$ | $^2P^\circ_{3/2}$ | 59463.494 | 72377.4616 | 1.89+7 | B+ | Brill | OH10 |
| | m | | | 7745.1(3) | | $5s^25f$ | $^2F^\circ_{5/2}$ | $5s^211d$ | $^2D_{3/2}$ | 99666.340 | 112574.1 | | | Wu | |
| 190 | | 7825.97(9) | 12774.45 | 7825.96(6) | 0.01 | $5s^27p$ | $^2P^\circ_{1/2}$ | $5s^29s$ | $^2S_{1/2}$ | 91903.958 | 104678.43 | 4.7+6 | D+ | Brill | TW |
| 280 | | 7903.532(4) | 12649.091 | 7903.531(3) | 0.001 | $5s5p^2$ | $^2D_{3/2}$ | $5s^26p$ | $^2P^\circ_{1/2}$ | 58844.194 | 71493.287 | 1.96+7 | B+ | Brill | OH10 |
| 53 | | 8055.72(9) | 12410.13 | 8055.60(6) | 0.12 | $5s^27p$ | $^2P^\circ_{3/2}$ | $5s^29s$ | $^2S_{1/2}$ | 92268.119 | 104678.43 | 8.6+6 | D+ | Brill | TW |
| 2600 | * | 9058.880(4) | 11035.863 | 9058.886(3) | −0.006 | $5s^24f$ | $^2F^\circ_{7/2}$ | $5s^25g$ | $^2G_{9/2}$ | 89288.268 | 100324.124 | 6.8+7 | D+ | Brill | TW |
| 2600 | * | 9058.880(4) | 11035.863 | 9058.886(3) | −0.006 | $5s^24f$ | $^2F^\circ_{7/2}$ | $5s^25g$ | $^2G_{7/2}$ | 89288.268 | 100324.124 | 2.4+6 | D+ | Brill | TW |
| 2200 | | 9063.658(5) | 11030.045 | 9063.649(4) | 0.009 | $5s^24f$ | $^2F^\circ_{5/2}$ | $5s^25g$ | $^2G_{7/2}$ | 89294.068 | 100324.124 | 6.5+7 | D+ | Brill | TW |
| 1300 | | 10607.434(6) | 9424.768 | 10607.430(6) | 0.004 | $5s^26d$ | $^2D_{3/2}$ | $5s^25f$ | $^2F^\circ_{5/2}$ | 90241.568 | 99666.340 | 4.9+7 | D+ | Brill | TW |
| 1200 | | 10739.257(6) | 9309.081 | 10739.255(5) | 0.002 | $5s^26d$ | $^2D_{5/2}$ | $5s^25f$ | $^2F^\circ_{7/2}$ | 90351.908 | 99660.991 | 5.1+7 | D+ | Brill | TW |
| | | | | 23521.08(23) | | $5s^25p$ | $^2P^\circ_{1/2}$ | $5s^25p$ | $^2P^\circ_{3/2}$ | 0.00 | 4251.505 | 6.94−1 M1 | A+ | | B95 |

[a] Observed relative intensities, in terms of total energy flux under the line profile, are reduced to a common arbitrary scale corresponding to a plasma in local thermodynamic equilibrium with an effective excitation temperature of 4.2 eV. These conditions correspond to exposure 1 of the experiment of Wu [11] (see section 4.4).

[b] Character of observed line: bl – blended by a close line (the blending spectrum is indicated in parentheses); h – hazy line; H – very hazy line; s – asymmetric line extending towards shorter wavelengths; * – intensity shared by two or more transitions; m – masked by a stronger neighboring line (no wavelength measured); : – the wavelength was not measured (the value in $\lambda_{obs}$ is a rounded Ritz wavelength).

[c] Observed and Ritz wavelengths are given in standard air for wavenumbers $\sigma$ between 5000 cm$^{-1}$ and 50000 cm$^{-1}$ and in vacuum outside of this range. The uncertainty (standard deviation) in the last digit is given in parentheses.

[d] Ritz wavelengths and their uncertainties were determined in the least-squares level optimization procedure (see section 4.3).

[e] Difference between observed and Ritz wavelength. If this column is blank, and $\lambda_{obs}$ is not blank, this line alone determines one of the levels involved in the assigned transition. An "x" after the value indicates that this line was excluded from the level optimization.

[f] In the transition probability values, the number after the "+" or "–" symbol means the power of 10.

[g] Transition probability accuracy code is explained in table 6.

[h] References to observed wavelengths and transition probabilities: AM – Alonso-Medina and Colón 2000 [34]; B95 – Biémont et al. 1995 [35]; Brill – Brill 1964 [11]; M79 – Miller et al. 1979 [39]; MS – McCormick and Sawyer 1938 [6]; OH10 – Oliver and Hibbert 2010 [14]; Wu – Wu 1967 [11]; Wu* – line measured by Wu [11] with our new or revised classification; TW – this work.

## 4. Results and discussion

### *4.1. Theoretical calculations*

The theoretical calculation for energy levels, wavelengths, and transition probabilities of Sn II was made with Cowan's codes [19], which implement the Hartree–Fock (HF) method with perturbative account for relativistic and configuration-interaction (CI) effects. For the even parity system, the configurations included were $5s5p^2$, $5s^2ns$ ($n = 6$–12), $5s^2nd$ ($n = 5$–12), $5s^2ng$ ($n = 5$–12), 4f5s5p, and $5s5d^2$; the odd parity set included $5s^2np$ ($n = 6$–20), $5s^2nf$ ($n = 4$–20), $5p^3$, 5s5p5d, 5s5p6s, 4f5s5d, $4f5p^2$, and $5p5d^2$ configurations. The initial scaling of the Slater parameters was 100 % of the HF values for $E_{av}$ and $\zeta_{nl}$, while the $F^k$, $G^k$, and the CI parameters were scaled to 80 % of the HF values. Then the Slater parameters were varied in the least-squares fitting (LSF) procedure minimizing discrepancies between calculated and observed level values.

In the parametric fitting, the energy level calculations for even parity converged with a standard deviation of 77 cm$^{-1}$, while the odd-parity configurations were fitted with a standard deviation of 156 cm$^{-1}$. Transition probabilities and autoionization rates were calculated with wavefunctions modified according to the fitted parameters.

### *4.2. Analysis of the spectrum*

#### *4.2.1. The $5s^25p$ – [$5s^2(ns+nd)$ + $5s5p^2$] transition array*

Excitation of the outer electron from the $5s^25p$ $^2P°$ ground term leads to the $5s^2ns$ $^2S_{1/2}$ and $5s^2nd$ $^2D_{3/2,5/2}$ level series showing a simple doublet structure. The transitions from $5s^2ns$ $^2S_{1/2}$ ($n = 6$–8) to both levels of the ground term and those from $5s^29s$ to $5s^25p$ $^2P°_{3/2}$ were already reported by McCormick and Sawyer [6]. The energy levels derived from their wavelengths were later improved by Shenstone and reported in AEL [4]. All these transitions are confirmed in our measurements with improved accuracy. McCormick and Sawyer [6] established the levels of $5s^210s$ and 11s by observing transitions to the $5s^26p$ levels in the air wavelength region. We were able to observe both transitions from $5s^29s$ to the levels of $5s^25p$ at 955.289 Å ($^2S_{1/2} \rightarrow {}^2P°_{1/2}$) and 995.738 Å ($^2S_{1/2} \rightarrow {}^2P°_{3/2}$). Wu [11] observed both transitions from $5s^210s$ to the ground-term levels. Other transitions from $5s^2ns$ ($n = 7$–11) to the $5s^2np$ ($n = 6$–7) levels have also been observed by us and by other researchers [6, 10, 11]. Thus, the levels of the $5s^2ns$ ($n = 6$–11) configurations are well established. They are the least perturbed, showing the leading *LS* percentages of their composition above 99 %. Our least-squares parametric fitting shows a good regularity and confirms all their identifications.

The $5s^2nd$ configurations were also listed in AEL [4]. We confirmed the levels of $5s^2nd$ ($n = 5$–9) with lines observed on our plates. Wu's identifications of transitions from the $5s^210d$ and 11d configurations [11] are also confirmed. Some additional transitions between the $5s^2np$ ($n = 6$–8) and $5s^2nd$ levels have also been observed (see table 1). It is important to mention here that there is a strong interaction between the $5s^25d$ and $5s5p^2$ configurations. For this reason, the $^2D_{3/2,5/2}$ levels of these configurations are strongly mixed each other. This strong mixing was indicated by relativistic CI calculations of Oliver and Hibbert [14], which, however, favored the old AEL designations. Percentage compositions of eigenvectors resulting from our calculations suggest that the configuration labels given in AEL for these two pairs of levels should be interchanged (see table 3). Nevertheless, to avoid confusion in line identifications we retained the AEL labels adopted also by Sansonetti and Martin [20].

Another type of excitation is represented by excitation of the inner 5s electron to the 5p shell, leading to the $5s5p^2$ configuration with seven levels containing a quartet and three doublet terms. For a long time it was difficult to observe the intercombination lines, as their transition probabilities are low. In the present work, we confirm six levels including $^2P_{1/2}$ at 80455.8 cm$^{-1}$ reported in AEL [4] as questionable. However, we could not confirm the $^2S_{1/2}$ level reported at 80206.1 cm$^{-1}$. This level value was ambiguous for two reasons. Firstly, it strongly deviated from the LSF calculations. Secondly, its strongest predicted transition to the ground level was missing. Therefore, this level value was rejected. We further disagree with the recent verification of this level value by Alonso-Medina et al. [17] since all transitions involving this level are very weak, and those observed by Alonso-Medina et al. [17] together with the claimed uncertainties cannot be reconciled with other identified lines in this spectrum. Connerade and Baig [21] revised the identification of the $5s5p^2$ $^2P_{1/2}$ and $^2P_{3/2}$ levels by analyzing level separations along the In I isoelectronic sequence. Their suggested values for these two levels were 80206 cm$^{-1}$ and 81718 cm$^{-1}$, respectively. We confirmed and refined the second level. However, the first one, as noted above, was found to be spurious. Calculations of Connerade and Baig [21] yielded a predicted value for the 5s5p$^2$ $^2S_{1/2}$ level at 60024 cm$^{-1}$. We located this level at a much higher position, at 75954.4 cm$^{-1}$, by identifying transitions from it to both levels of the ground term. The strongest of these transitions (to $^2P_{1/2}$) was observed in our spectra at 1316.579 Å. The transition to $^2P_{3/2}$, predicted to be much weaker, was not observed on our plates. However, it was observed by Wu [11] at 1394.667 Å in two exposures. This newly revised level value fits well in our parametric LSF calculations with reasonable values of the fitted parameters. This identification is further validated by an isoelectronic comparison presented in figure 2 for the sequence In I to Xe VI. The $^2S_{1/2}$ and $^2P_{1/2}$ levels of 5s5p$^2$ are strongly mixed in these spectra. In In I, the leading terms are $^2$S and $^2$P for the lower and upper of these two levels, respectively, while in Xe VI they are reversed. The currently adopted position of 5s5p$^2$ $^2S_{1/2}$ level in Sn II, indicated by dashed lines in figure 2, is strikingly inconsistent with the smooth isoelectronic trend of other data points. Our new LSF calculations for this sequence result in interchange of the term labels $^2S_{1/2}$ and $^2P_{1/2}$ in Te IV and predict a much lower position for the $^2S_{1/2}$ level in Sn II. This prediction is in qualitative agreement with findings of Connerade and Baig [21]. As indicated by the solid lines in figure 2, our revised identification produces a smooth isoelectronic trend for the lower $J = 1/2$ level, similar to the behavior of the upper level. The revised level values and term labels, along with the calculated percentage compositions, are given in table 2. Additional support for our new identification of the 5s5p$^2$ $^2S_{1/2}$ level in Sn II is provided by a recent theoretical calculation by Oliver and Hibbert [14]. They predicted $^2S_{1/2}$ in Sn II to be at 76215 cm$^{-1}$, which is in close agreement with our newly found level value. Colón and Alonso-Medina [22] suggested an explanation of the anomaly in the 5s5p$^2$ $^2S_{1/2}$ and $^2P_{1/2}$ levels of Sn II by the presence of some mysterious interacting configuration(s). As this anomaly is now resolved, their suggestion can be dismissed. It should be noted that these two levels are strongly mixed (see tables 2 and 3). Thus, their *LS* labels have little physical meaning and are used in our tables for bookkeeping purpose only.

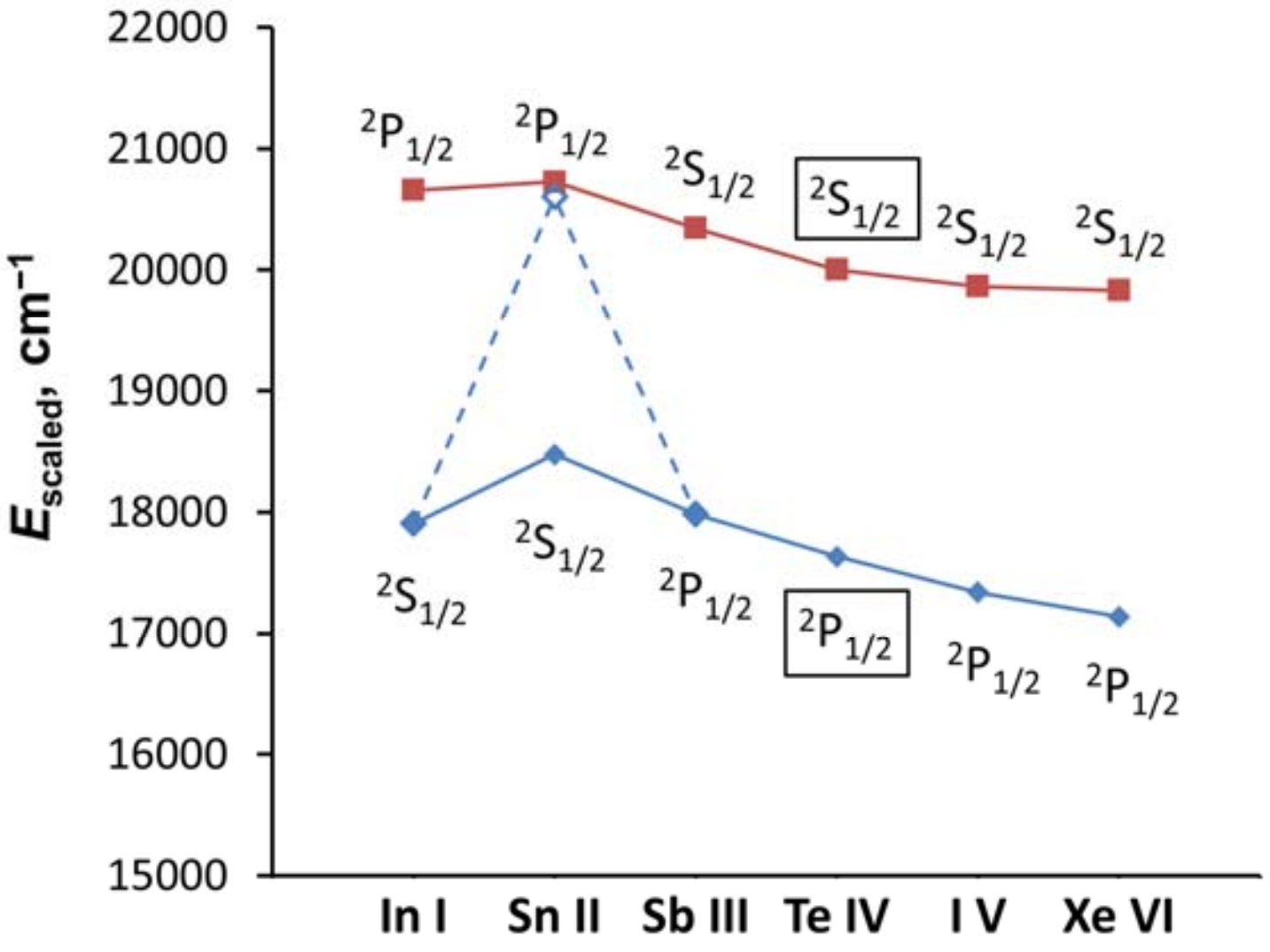

Figure 2 (*color online*). Isoelectronic comparison of scaled energies, $E_{scaled} = (E - 39000)/Z_c$, of the strongly mixed 5s5p$^2$ $^2S_{1/2}$ and $^2P_{1/2}$ levels. The dominant term labels of the lower and upper levels interchange at ionic core charge $Z_c = 3$ (Sb III). The open circle just below the $^2P_{1/2}$ data point for Sn II indicates the previously adopted position of the $^2S_{1/2}$ level in Sn II at 80206 cm$^{-1}$ [4]. Dashed lines connecting this data point with the other ones of the lower level show how this graph would look if that incorrect value were used instead of our revised value (solid rhomb and solid lies). Solid boxes indicate the revised term labels for Te IV. See table 2 for details.

Table 2. The two $J = 1/2$ doublet levels of the $5s5p^2$ configuration in the In I isoelectronic sequence. All percentage compositions and the Sn II energies are from the present work; the rest of the data are from ASD [5].

| Isoelectronic member | Lower level | | | | Upper level | | | |
|---|---|---|---|---|---|---|---|---|
| | Energy, $cm^{-1}$ | Percentages | | Term label | Energy, $cm^{-1}$ | Percentage | | Term label |
| | | $^2S$ | $^2P$ | | | $^2S$ | $^2P$ | |
| In I | 56906 | 74 | 24 | $^2S$ | 59657 | 23 | 76 | $^2P$ |
| Sn II | 75954.4 [a] | 51 | 46 | $^2S$ | 80455.3 | 45 | 53 | $^2P$ |
| Sb III | 92948.9 | 31 | 37 | $^2P$ [b] | 100030 | 57 | 41 | $^2S$ |
| Te IV | 109536 | 40 | 57 | $^2P$ [c] | 119009 | 56 | 42 | $^2S$ [c] |
| I V | 125703 | 36 | 61 | $^2P$ | 138328 | 60 | 38 | $^2S$ |
| Xe VI | 141837 | 32 | 64 | $^2P$ | 157995 | 62 | 36 | $^2S$ |

[a] Our revised value replaces the previously reported 80206 $cm^{-1}$ [4].
[b] Third leading component: 31 % of $5s^26s\ ^2S$ (at 93422.5 $cm^{-1}$).
[c] The level designations for Te IV are interchanged according to our LSF calculations.

*4.2.2. The $5p^3$ and 5s5p(5d+6s) configurations*

These configurations arise from excitation of the $5s5p^2$ configuration. In the sequence Sb III – I V [23–25], transitions from these configurations have been observed in the Antigonish laboratory with moderate intensity. Therefore, we expected them to occur in Sn II as well. Our preliminary calculations for Sn II show that these configurations strongly interact with each other and also with other configurations, e.g., 4f5s5d and $5p5d^2$, which are completely unknown at present. The $5p^3$ and 5s5p(5d+6s) configurations are predicted to extend past the ionization limit. Thus, many of their levels should be autoionizing, making the analysis more difficult. Additional complication stems from the fact that some of the levels of these configurations are embedded within highly excited levels of the $5s^2n$p and $n$f series, with which they strongly interact. A few levels of these configurations were listed in AEL [4] with incomplete designations; some were marked as doubtful. We attempted to improve interpretation of these levels. The level at 109223.4 $cm^{-1}$ in AEL [4] is supported by two transitions terminating on the $5s5p^2\ ^4P_{1/2,3/2}$ levels, observed in our spectra, and one transition to $5s^25d\ ^2D_{3/2}$ observed by Wu [11]. We now identified this level as $5s^210p\ ^2P_{1/2}$ on the basis of our LSF calculation. Observed relative intensities of the lines are in satisfactory agreement with calculations. We were able to confirm the quartet levels of 5s5p6s configuration listed in AEL [4], as they give rise to strong transitions to the lowest quartet levels of the $5s5p^2$ configuration. The level at 113819.0 $cm^{-1}$ is also confirmed. In AEL [4], the $J$ value of this level was given as 3/2 with a question mark, and no configuration label was attributed to it. Our present calculation with extensive configuration interaction shows that this level should be assigned to $^2D_{5/2}$ of the 5s5p5d configuration.

As noted above, identification of autoionizing levels presented considerable difficulties. We could not confirm the level at 124627.7 $cm^{-1}$ in AEL [4]. Wu [11] assigned three observed lines at 1520.153 Å, 2907.33 Å, and 7412.5 Å to this level. However, no satisfactory match could be found for this level in our calculations. All other autoionizing levels given in AEL [4] have been identified in our spectra. In particular, three levels previously reported at 123156, 132168 and 132708 $cm^{-1}$ with uncertain designations are found to be associated with the $5p^3$ configuration. The first of them is identified as $5p^3\ ^4S°_{3/2}$, while the other two have dominant contributions of $5s5p(^1P°)5d\ ^2D°_{3/2}$ and $^2D°_{5/2}$, respectively. Since the 5s5p5d configuration strongly interacts with $5p^3$, these levels have large admixtures of $5p^3\ ^2D°$ in their wavefunctions. A few of the other autoionizing levels that were based on just one or two observed transitions remain questionable.

*4.2.3. The $5s^2n$p and $5s^2n$f configurations*

After the successful establishment of the $5s^2n$s and $n$d levels, a further analysis of the $5s^2n$p and $n$f configurations was undertaken. The $5s^2n$p ($n$ = 6–9) and $5s^2n$f ($n$ = 4–6) configurations were already reported in AEL [4]. Some transitions from $5s^26$p and 7p to levels of the $5s^26$s, $5s^25$d, and $5s5p^2$ configurations were measured interferometrically by Brill [10]. Lines arising from the $5s^2n$p ($n$ = 6–9) configurations were classified by McCormick and Sawyer [6], and some additional lines were also observed by Wu [11]. We confirm all these identifications, as the observed level energies and relative line intensities are in satisfactory agreement with our calculations.

#### 4.2.4. Levels of $5s^2ng$ configurations

The $5s^2ng$ ($n = 6$–11) $^2G$ energy levels were established by transitions from the levels of $5s^24f$ configuration, identified primarily by McCormick and Sawyer [6]. Some of the transitions observed by McCormick and Sawyer were more accurately measured by Wu [11]. Brill [10] re-measured the 4f–6g transitions with much better accuracy. No discernible fine-structure splitting was detected in any of the observed $5s^2ng$ $^2G$ multiplets. The lowest member of this series, $5s^25g$ $^2G$, was unknown so far. Brill [10] observed a pair of lines at 9058.880 Å and 9063.658 Å with a separation 5.818(6) cm$^{-1}$ closely matching his measured $5s^24f$ $^2F°$ $J = 5/2$–$7/2$ interval, 5.804(9) cm$^{-1}$. He recognized that these lines must correspond to transitions combining the $5s^24f$ $^2F°_{5/2,7/2}$ levels with some unknown level, but he was unable to decide whether this level is located above or below $5s^24f$ $^2F°$. Thus, he gave two possible energy values for this unknown level, 78258.194(6) cm$^{-1}$ or 100324.103(6) cm$^{-1}$. By extrapolating the known energies of the $5s^2ng$ $^2G$ terms with $n \geq 6$ to $n = 5$ with the core-polarization formula (see section 4.5), we found that the upper of these two suggested values almost exactly coincides with the predicted position of the $5s^25g$ $^2G$ term. Our LSF calculations ruled out the existence of a level at the lower of the two positions suggested by Brill, that could possibly combine with $5s^24f$ $^2F°$. Thus, we identified the level at 100324.103(6) cm$^{-1}$ as $5s^25g$ $^2G$. Observed level energies and relative line intensities of all transitions from $5s^2ng$ levels agree well with our calculations.

### 4.3. Optimization of energy levels

To derive the energy level values that best fit all observed transition wavelengths, we used the least-squares level optimization program LOPT [26]. The crucial factors for the level optimization procedure are the correct identification of the spectral lines, estimation of their uncertainties, and absence of systematic shifts. Correctness of identifications was ensured by the analysis described above. Estimation of the statistical and systematic uncertainties of wavelengths was described in section 3. This estimation partially relies on the level optimization procedure, since some of the reference wavelengths used in this procedure are the Sn II Ritz wavelengths. Therefore, the level optimization was made in several iterations. In the initial stage, only the accurate measurements of Brill [10], as well as our measurements in the VUV, for which independent estimates of uncertainties are available, were included in the optimization. This resulted in initial estimates of the energy levels and Ritz wavelengths derived from them. Deviations of wavelengths observed by Wu [11] and by McCormick and Sawyer [6] from these Ritz wavelengths revealed systematic shifts smoothly varying with wavelength. After these systematic shifts were removed, residual deviations of corrected wavelengths from Ritz values provided a sufficient statistical basis to assign uncertainties to all the measurements. Then the corrected wavelengths from Wu [11] and McCormick and Sawyer [6] were also included in the level optimization, leading to an extended and more accurate set of energy levels and Ritz wavelengths. This process was repeated until the estimated systematic shifts stopped changing.

Table 3. Optimized energy levels of Sn II

| Configuration | Term | $J$ | Energy [a] cm$^{-1}$ | | Unc. [b] cm$^{-1}$ | Leading percentages [c] | | | | $\Delta E_{o-c}$ [d] | No. of lines [e] |
|---|---|---|---|---|---|---|---|---|---|---|---|
| $5s^25p$ | $^2P°$ | 1/2 | 0.00 | | 0.04 | 97 | 2 | $5p^3$ | $^2P°$ | 81 | 17 |
| $5s^25p$ | $^2P°$ | 3/2 | 4251.505 | | 0.014 | 97 | 2 | $5p^3$ | $^2P°$ | -81 | 25 |
| $5s5p^2$ | $^4P$ | 1/2 | 46464.301 | | 0.018 | 98 | | | | -63 | 12 |
| $5s5p^2$ | $^4P$ | 3/2 | 48368.198 | | 0.007 | 99 | | | | -20 | 17 |
| $5s5p^2$ | $^4P$ | 5/2 | 50730.237 | | 0.005 | 97 | 3 | $5s5p^2$ | $^2D$ | 80 | 12 |
| $5s^26s$ | $^2S$ | 1/2 | 56886.3763 | | 0.003 | 100 | | | | 0 | 7 |
| $5s5p^2$ | $^2D$ | 3/2 | 58844.194 | | 0.004 | 41 | 58 | $5s^25d$ | $^2D$ | -128 | 17 |
| $5s5p^2$ | $^2D$ | 5/2 | 59463.494 | | 0.005 | 38 | 59 | $5s^25d$ | $^2D$ | 128 | 12 |
| $5s^25d$ | $^2D$ | 3/2 | 71406.155 | | 0.008 | 41 | 54 | $5s5p^2$ | $^2D$ | -120 | 11 |
| $5s^26p$ | $^2P°$ | 1/2 | 71493.287 | | 0.003 | 99 | | | | 3 | 14 |
| $5s^25d$ | $^2D$ | 5/2 | 72048.273 | | 0.007 | 40 | 56 | $5s5p^2$ | $^2D$ | 119 | 12 |
| $5s^26p$ | $^2P°$ | 3/2 | 72377.4616 | | | 99 | | | | -3 | 16 |
| $5s5p^2$ | $^2S$ | 1/2 | 75954.4 | R | 0.3 | 51 | 46 | $5s5p^2$ | $^2P$ | 105 | 3 |
| $5s5p^2$ | $^2P$ | 1/2 | 80455.3 | C | 0.3 | 53 | 45 | $5s5p^2$ | $^2S$ | -120 | 3 |
| $5s5p^2$ | $^2P$ | 3/2 | 81718.43 | | 0.24 | 97 | 2 | $5s5p^2$ | $^2D$ | 18 | 3 |
| $5s^27s$ | $^2S$ | 1/2 | 86280.332 | | 0.005 | 99 | | | | 0 | 5 |
| $5s^24f$ | $^2F°$ | 7/2 | 89288.268 | | 0.007 | 96 | 3 | $4f5p^2$ | $^2F°$ | -1 | 10 |
| $5s^24f$ | $^2F°$ | 5/2 | 89294.068 | | 0.006 | 96 | 3 | $4f5p^2$ | $^2F°$ | 1 | 11 |

| Configuration | Term | $J$ | Energy [a] cm$^{-1}$ | | Unc. [b] cm$^{-1}$ | Leading percentages [c] | | | | $\Delta E_{o-c}$ [d] | No. of lines [e] |
|---|---|---|---|---|---|---|---|---|---|---|---|
| 5s$^2$6d | $^2$D | 3/2 | 90241.568 | | 0.004 | 97 | 2 | 5s5p$^2$ | $^2$D | -27 | 7 |
| 5s$^2$6d | $^2$D | 5/2 | 90351.908 | | 0.005 | 97 | 2 | 5s5p$^2$ | $^2$D | 26 | 6 |
| 5s$^2$7p | $^2$P° | 1/2 | 91903.958 | | 0.015 | 99 | | | | -2 | 6 |
| 5s$^2$7p | $^2$P° | 3/2 | 92268.119 | | 0.009 | 99 | | | | 2 | 10 |
| 5s$^2$8s | $^2$S | 1/2 | 98402.425 | | 0.006 | 100 | | | | 0 | 5 |
| 5s$^2$5f | $^2$F° | 7/2 | 99660.991 | | 0.006 | 99 | | | | -1 | 4 |
| 5s$^2$5f | $^2$F° | 5/2 | 99666.340 | | 0.006 | 99 | | | | 1 | 4 |
| 5s$^2$7d | $^2$D | 3/2 | 100284.125 | | 0.010 | 99 | | | | -11 | 6 |
| 5s$^2$5g | $^2$G | 7/2 | 100324.124 | C | 0.007 | 100 | | | | 0 | 2 |
| 5s$^2$5g | $^2$G | 9/2 | 100324.124 | C | 0.007 | 100 | | | | 0 | 1 |
| 5s$^2$7d | $^2$D | 5/2 | 100338.960 | | 0.009 | 99 | | | | 11 | 2 |
| 5s$^2$8p | $^2$P° | 1/2 | 101204.2 | | 0.4 | 99 | | | | 4 | 2 |
| 5s$^2$8p | $^2$P° | 3/2 | 101387.18 | | 0.21 | 99 | | | | 1 | 5 |
| 5s$^2$9s | $^2$S | 1/2 | 104678.43 | | 0.10 | 100 | | | | 0 | 6 |
| 5s$^2$6f | $^2$F° | 7/2 | 105360.3 | | 1.8 | 99 | | | | 9 | 1 |
| 5s$^2$6f | $^2$F° | 5/2 | 105371.7 | N | 0.3 | 99 | | | | 17 | 2 |
| 5s$^2$8d | $^2$D | 3/2 | 105731.2 | | 0.3 | 100 | | | | -6 | 5 |
| 5s$^2$6g | $^2$G | 7/2 | 105737.495 | | 0.007 | 100 | | | | 0 | 2 |
| 5s$^2$6g | $^2$G | 9/2 | 105737.495 | | 0.007 | 100 | | | | 0 | 1 |
| 5s$^2$8d | $^2$D | 5/2 | 105763.02 | | 0.23 | 100 | | | | 6 | 3 |
| 5s$^2$9p | $^2$P° | 1/2 | 106256.4 | N | 0.3 | 99 | | | | -28 | 2 |
| 5s$^2$9p | $^2$P° | 3/2 | 106370.2 | | 0.4 | 99 | | | | -26 | 3 |
| 5s$^2$10s | $^2$S | 1/2 | 108357.6 | | 0.4 | 100 | | | | 0 | 4 |
| 5s$^2$7g | $^2$G | 7/2 | 109002.3 | | 0.4 | 100 | | | | 0 | 2 |
| 5s$^2$7g | $^2$G | 9/2 | 109002.3 | | 0.4 | 100 | | | | 0 | 1 |
| 5s$^2$9d | $^2$D | 3/2 | 109006.2 | | 0.4 | 100 | | | | -4 | 3 |
| 5s$^2$9d | $^2$D | 5/2 | 109025.68 | | 0.15 | 100 | | | | 4 | 4 |
| 5s$^2$10p | $^2$P° | 1/2 | 109222.18 | R | 0.23 | 81 | 16 | 5s5p($^3$P°)6s | $^4$P° | 61 | 3 |
| 5s5p($^3$P°)6s | $^4$P° | 1/2 | 109454.05 | | 0.18 | 81 | 18 | 5s$^2$10p | $^2$P° | -13 | 2 |
| 5s$^2$11s | $^2$S | 1/2 | 110699 | | 4 | 100 | | | | 1 | 1 |
| 5s5p($^3$P°)6s | $^4$P° | 3/2 | 110777.85 | | 0.12 | 79 | 12 | 5s$^2$11p | $^2$P° | -11 | 5 |
| 5s$^2$8g | $^2$G | 7/2 | 111120.8 | | 0.4 | 100 | | | | 0 | 2 |
| 5s$^2$8g | $^2$G | 9/2 | 111120.8 | | 0.4 | 100 | | | | 0 | 1 |
| 5s$^2$10d | $^2$D | 3/2 | 111122.6 | | 1.2 | 100 | | | | -7 | 2 |
| 5s$^2$10d | $^2$D | 5/2 | 111143.3 | | 1.1 | 100 | | | | 5 | 4 |
| 5s$^2$9g | $^2$G | 7/2 | 112574.0 | | 0.8 | 100 | | | | 0 | 2 |
| 5s$^2$9g | $^2$G | 9/2 | 112574.0 | | 0.8 | 100 | | | | 0 | 1 |
| 5s$^2$11d | $^2$D | 3/2 | 112574.1 | N | 0.5 | 100 | | | | -8 | 2 |
| 5s$^2$11d | $^2$D | 5/2 | 112596.9 | | 1.1 | 100 | | | | 9 | 3 |
| 5s$^2$10g | $^2$G | 7/2 | 113610.5 | | 1.5 | 100 | | | | 0 | 2 |
| 5s$^2$10g | $^2$G | 9/2 | 113610.5 | | 1.5 | 100 | | | | 0 | 1 |
| 5s5p($^3$P°)5d | $^2$D° | 5/2 | 113818.68 | R | 0.18 | 62 | 18 | 5p$^3$ | $^2$D° | 44 | 4 |
| 5s5p($^3$P°)6s | $^2$P° | 1/2 | 113952.66 | N | 0.13 | 30 | 50 | 5s$^2$14p | $^2$P° | 9 | 6 |
| 5s5p($^3$P°)6s | $^4$P° | 5/2 | 114245.98 | | 0.14 | 95 | 2 | 5s5p($^3$P°)5d | $^2$D° | 9 | 6 |
| 5s$^2$11g | $^2$G | 7/2,9/2 | 114378.1 | | 2.2 | 100 | | | | 0 | 1 |
| 5s$^2$11g | $^2$G | 7/2,9/2 | 114378.1 | | 2.2 | 100 | | | | 0 | 1 |
| 5s5p($^3$P°)5d | $^4$F° | 5/2 | 115256.35 | N? | 0.22 | 76 | 8 | 5s5p($^3$P°)5d | $^2$D° | -31 | 1 |
| 5s5p($^3$P°)5d | $^4$F° | 7/2 | 116423.4 | N | 0.3 | 50 | 41 | 5s$^2$19f | $^2$F° | -49 | 3 |
| Sn III 5s$^2$ $^1$S$_0$ | Limit | | **118023.7** | R | 0.5 | | | | | | |
| 5s5p($^3$P°)5d | $^4$D° | 1/2 | 119981.3 | R? | 0.4 | 93 | 4 | 5s5p($^3$P°)5d | $^4$P° | -116 | 2 |
| 5s5p($^3$P°)5d | $^4$D° | 3/2 | 120064.5 | R? | 0.4 | 76 | 16 | 5s5p($^3$P°)5d | $^4$P° | -73 | 2 |
| 5s5p($^3$P°)5d | $^4$D° | 5/2 | 120253.85 | R | 0.23 | 52 | 39 | 5s5p($^3$P°)5d | $^4$P° | 54 | 3 |
| 5s5p($^3$P°)5d | $^4$D° | 7/2 | 122491.6 | R? | 0.5 | 92 | 7 | 5s5p($^3$P°)5d | $^4$F° | 102 | 1 |
| 5p$^3$ | $^4$S° | 3/2 | 123156.8 | R | 0.3 | 88 | 5 | 5p$^3$ | $^2$P° | -2 | 4 |
| 5s5p($^3$P°)5d | $^4$P° | 5/2 | 123688.3 | N | 0.4 | 55 | 41 | 5s5p($^3$P°)5d | $^4$D° | 60 | 2 |
| 5s5p($^3$P°)5d | $^4$P° | 3/2 | 124245.80 | R | 0.16 | 80 | 18 | 5s5p($^3$P°)5d | $^4$D° | 15 | 8 |
| 5s5p($^3$P°)5d | $^4$P° | 1/2 | 124524.4 | N? | 0.6 | 94 | 5 | 5s5p($^3$P°)5d | $^4$D° | -13 | 1 |

| Configuration | Term | $J$ | Energy [a] $cm^{-1}$ | | Unc. [b] $cm^{-1}$ | Leading percentages [c] | | | | $\Delta E_{o-c}$ [d] | No. of lines [e] |
|---|---|---|---|---|---|---|---|---|---|---|---|
| 5s5p($^1$P°)5d | $^2$D° | 3/2 | 132168.95 | R | 0.17 | 51 | 32 | 5p$^3$ | $^2$D° | -382 | 4 |
| 5s5p($^1$P°)5d | $^2$D° | 5/2 | 132708.1 | R | 0.3 | 54 | 32 | 5p$^3$ | $^2$D° | 365 | 4 |

[a] Symbols next to the energy value have the following meaning: C – previous tentative identification has been confirmed here; N – new identification; R – previous value and/or interpretation has been revised here; ? – questionable identification.

[b] Uncertainties resulting from the level optimization procedure are given on the level of one standard deviation. They correspond to uncertainties of level separations from 5s$^2$6p $^2$P°$_{3/2}$. To determine uncertainties of excitation energies from the ground level, the given values should be combined in quadrature with the uncertainty of the ground level, 0.04 $cm^{-1}$.

[c] The first percentage value refers to the configuration and term given in the first two columns of the table. The second percentage value refers to the configuration and term given next to it. The percentage compositions were determined in this work by a parametric least-squares fitting with Cowan's codes [19] (see text).

[d] Differences between observed energies and those calculated in the parametric least squares fitting.

[e] Number of observed lines determining the level in the optimization procedure.

The final list of optimized energy levels is given in table 3. In this table, the level uncertainties are given for separations from the 5s$^2$6p $^2$P°$_{3/2}$ level. This level was chosen as the base, because it has the largest number of accurately measured transitions. To infer the uncertainty of an excitation energy from the ground level, one should combine the given uncertainty value in quadrature with the uncertainty of the ground level, 0.04 $cm^{-1}$. It can be seen that the level uncertainties vary greatly, from 0.003 $cm^{-1}$ to 4 $cm^{-1}$, depending on the number and measurement accuracy of the lines determining the level. With a few exceptions, the level values are rounded using the "rule of 24," i.e., the uncertainty of the value does not exceed 24 units of the least significant digit of the value. In a few cases, an additional significant figure was required in order to reproduce the precisely measured transition wavelengths.

Some of the results of our LSF calculations, such as percentage compositions and differences of observed energies from those calculated in the parametric fitting are also given in table 3. The fitted parameter values obtained in the LSF are given in table 4.

Table 4. LSF parameters ($cm^{-1}$) for Sn II

| Configuration | Parameter | LSF | Group [a] | STD | HFR | LSF/HFR |
|---|---|---|---|---|---|---|
| **Odd Parity [b]** | | | | | | |
| 5s$^2$5p | $E_{av}$ | 6860.3 | | 116 | 0.0 | |
| | $\zeta$(5p) | 3016.3 | 1 | 83 | 2665.8 | 1.1315 |
| 5s$^2$6p | $E_{av}$ | 72666.4 | | 111 | 70522.8 | 1.0304 |
| | $\zeta$(6p) | 610.5 | 1 | 17 | 539.6 | 1.1314 |
| 5s$^2$7p | $E_{av}$ | 92361.0 | | 111 | 90536.8 | 1.0201 |
| | $\zeta$(7p) | 246.7 | 1 | 7 | 218.0 | 1.132 |
| 5s$^2$8p | $E_{av}$ | 101458.0 | | 111 | 99608.5 | 1.0186 |
| | $\zeta$(8p) | 127.0 | 1 | 3 | 112.2 | 1.132 |
| 5s$^2$9p | $E_{av}$ | 106459.7 | 4 | 99 | 104563.0 | 1.0181 |
| | $\zeta$(9p) | 74.2 | 1 | 2 | 65.6 | 1.132 |
| 5s$^2$10p | $E_{av}$ | 109339.8 | 4 | 101 | 107572.2 | 1.0164 |
| | $\zeta$(10p) | 47.2 | 1 | 1 | 41.7 | 1.132 |
| 5s$^2$11p | $E_{av}$ | 111223.6 | 4 | 103 | 109540.5 | 1.0154 |
| | $\zeta$(11p) | 31.8 | 1 | 1 | 28.1 | 1.132 |
| 5s$^2$12p | $E_{av}$ | 112521.2 | 4 | 104 | 110896.3 | 1.0147 |
| | $\zeta$(12p) | 22.5 | 1 | 1 | 19.9 | 1.131 |
| 5s$^2$13p | $E_{av}$ | 113454.3 | 4 | 105 | 111871.2 | 1.0142 |

| Configuration | Parameter | LSF | Group [a] | STD | HFR | LSF/HFR |
|---|---|---|---|---|---|---|
| | $\zeta$(13p) | 16.4 | 1 | 0 | 14.5 | 1.131 |
| $5s^2$14p | $E_{av}$ | 114151.8 | 4 | 106 | 112600.0 | 1.0138 |
| | $\zeta$(14p) | 12.4 | 1 | 0 | 11.0 | 1.13 |
| $5s^2$15p | $E_{av}$ | 114680.1 | 4 | 106 | 113152.0 | 1.0135 |
| | $\zeta$(15p) | 9.6 | 1 | 0 | 8.5 | 1.13 |
| $5s^2$16p | $E_{av}$ | 115095.4 | 4 | 107 | 113585.9 | 1.0133 |
| | $\zeta$(16p) | 7.6 | 1 | 0 | 6.7 | 1.13 |
| $5s^2$17p | $E_{av}$ | 115427.6 | 4 | 107 | 113933.0 | 1.0131 |
| | $\zeta$(17p) | 6.0 | 1 | 0 | 5.3 | 1.13 |
| $5s^2$18p | $E_{av}$ | 115694.7 | 4 | 107 | 114212.1 | 1.0130 |
| | $\zeta$(18p) | 4.9 | 1 | 0 | 4.3 | 1.14 |
| $5s^2$19p | $E_{av}$ | 115910.3 | 4 | 108 | 114437.3 | 1.0129 |
| | $\zeta$(19p) | 4.1 | 1 | 0 | 3.6 | 1.14 |
| $5s^2$20p | $E_{av}$ | 116095.8 | 4 | 108 | 114631.2 | 1.0128 |
| | $\zeta$(20p) | 3.4 | 1 | 0 | 3.0 | 1.13 |
| $4f5s^2$ | $E_{av}$ | 93783.8 | | 114 | 87380.8 | 1.0733 |
| | $\zeta$(4f) | 0.4 | | fixed | 0.4 | 1.0000 |
| $5s^2$5f | $E_{av}$ | 99981.2 | | 112 | 97859.7 | 1.0217 |
| | $\zeta$(5f) | 0.2 | | fixed | 0.2 | 1.0000 |
| $5s^2$6f | $E_{av}$ | 105538.7 | 5 | 107 | 103503.1 | 1.0197 |
| | $\zeta$(6f) | 0.1 | | fixed | 0.1 | 1.0000 |
| $5s^2$7f | $E_{av}$ | 108779.9 | 5 | 110 | 106886.6 | 1.0177 |
| | $\zeta$(7f) | 0.1 | | fixed | 0.1 | 1.0000 |
| $5s^2$8f | $E_{av}$ | 110873.6 | 5 | 112 | 109072.3 | 1.0165 |
| | $\zeta$(8f) | 0.1 | | fixed | 0.1 | 1.0000 |
| $5s^2$9f | $E_{av}$ | 112303.7 | 5 | 114 | 110565.2 | 1.0157 |
| $5s^2$10f | $E_{av}$ | 113325.1 | 5 | 115 | 111631.4 | 1.0152 |
| $5s^2$11f | $E_{av}$ | 114076.2 | 5 | 116 | 112415.5 | 1.0148 |
| $5s^2$12f | $E_{av}$ | 114646.3 | 5 | 116 | 113010.6 | 1.0145 |
| $5s^2$13f | $E_{av}$ | 115090.9 | 5 | 117 | 113474.8 | 1.0142 |
| $5s^2$14f | $E_{av}$ | 115440.1 | 5 | 117 | 113839.3 | 1.0141 |
| $5s^2$15f | $E_{av}$ | 115719.7 | 5 | 117 | 114131.2 | 1.0139 |
| $5s^2$16f | $E_{av}$ | 115953.9 | 5 | 118 | 114375.6 | 1.0138 |
| $5s^2$17f | $E_{av}$ | 116144.1 | 5 | 118 | 114574.2 | 1.0137 |
| $5s^2$18f | $E_{av}$ | 116301.7 | 5 | 118 | 114738.7 | 1.0136 |
| $5s^2$19f | $E_{av}$ | 116444.8 | 5 | 118 | 114888.1 | 1.0135 |
| $5s^2$20f | $E_{av}$ | 116564.1 | 5 | 118 | 115012.6 | 1.0135 |
| 5s5p6s | $E_{av}$ | 118905.2 | | 211 | 109900.0 | 1.0819 |
| | $\zeta$(5p) | 3492.9 | 1 | 96 | 3087.0 | 1.1315 |
| | $G^1$(5s5p) | 29907.8 | 6 | 454 | 51164.2 | 0.5845 |
| | $G^0$(5s6s) | 1745.8 | 9 | 246 | 2522.8 | 0.6920 |
| | $G^1$(5p6s) | 2925.9 | 9 | 413 | 4228.1 | 0.6920 |
| 5s5p5d | $E_{av}$ | 128129.4 | | 152 | 117237.6 | 1.0929 |
| | $\zeta$(5p) | 3415.7 | 1 | 94 | 3018.8 | 1.1315 |

| Configuration | Parameter | LSF | Group [a] | STD | HFR | LSF/HFR |
|---|---|---|---|---|---|---|
| | $\zeta$(5d) | 77.2 | | fixed | 77.2 | 1.0000 |
| | $F^2$(5p5d) | 20634.2 | | 865 | 20388.4 | 1.0121 |
| | $G^1$(5s5p) | 29712.1 | 6 | 451 | 50829.4 | 0.5845 |
| | $G^2$(5s5d) | 10985.9 | | 1189 | 9761.6 | 1.1254 |
| | $G^1$(5p5d) | 18655.8 | 7 | 663 | 20189.5 | 0.9240 |
| | $G^3$(5p5d) | 11538.5 | 7 | 410 | 12487.0 | 0.9240 |
| 5p$^3$ | $E_{av}$ | 136589.5 | | 179 | 127058.9 | 1.0750 |
| | $F^2$(5p5p) | 32281.3 | | fixed | 37046.5 | 0.8714 |
| | $\zeta$(5p) | 3064.7 | 1 | 84 | 2708.6 | 1.1315 |
| 5p5d$^2$ [c] | $E_{av}$ | 262609.7 | | fixed | 255939.8 | 1.0261 [c] |
| | $\zeta$(5p) | 3732.4 | 1 | 103 | 3298.7 | 1.1315 |
| 4f5p$^2$ [c] | $E_{av}$ | 220238.9 | | fixed | 213569.0 | 1.0312 [c] |
| | $\zeta$(5p) | 3621.3 | 1 | 100 | 3200.5 | 1.1315 |
| 4f5s5d [c] | $E_{av}$ | 223939.3 | | fixed | 217269.4 | 1.0307 [c] |
| **Even Parity** | | | | | | |
| 5s5p$^2$ | $E_{av}$ | 63956.0 | | 36 | 55474.7 | 1.1529 |
| | $F^2$(5p5p) | 32090.9 | | 286 | 36826.8 | 0.8714 |
| | $\zeta$(5p) | 3040.9 | | 57 | 2681.9 | 1.1339 |
| | $G^1$(5s5p) | 30953.3 | | 109 | 48632.5 | 0.6365 |
| 5s$^2$6s | $E_{av}$ | 56981.0 | | 77 | 55518.4 | 1.0263 |
| 5s$^2$7s | $E_{av}$ | 86243.4 | | 77 | 84679.3 | 1.0185 |
| 5s$^2$8s | $E_{av}$ | 98398.8 | | 77 | 96666.1 | 1.0179 |
| 5s$^2$9s | $E_{av}$ | 104677.5 | | 77 | 102872.0 | 1.0176 |
| 5s$^2$10s | $E_{av}$ | 108357.2 | | 77 | 106511.0 | 1.0173 |
| 5s$^2$11s | $E_{av}$ | 110698.2 | 2 | 77 | 108830.1 | 1.0172 |
| 5s$^2$12s | $E_{av}$ | 112202.1 | 2 | 78 | 110399.3 | 1.0163 |
| 5s$^2$5d | $E_{av}$ | 64921.4 | | 100 | 64114.7 | 1.0126 |
| | $\zeta$(5d) | 66.7 | | fixed | 66.7 | 1.0000 |
| 5s$^2$6d | $E_{av}$ | 89806.5 | | 56 | 88201.0 | 1.0182 |
| | $\zeta$(6d) | 26.7 | | fixed | 26.7 | 1.0000 |
| 5s$^2$7d | $E_{av}$ | 100165.1 | | 55 | 98423.5 | 1.0177 |
| | $\zeta$(7d) | 13.7 | | fixed | 13.7 | 1.0000 |
| 5s$^2$8d | $E_{av}$ | 105680.4 | | 54 | 103874.8 | 1.0174 |
| | $\zeta$(8d) | 7.9 | | fixed | 7.9 | 1.0000 |
| 5s$^2$9d | $E_{av}$ | 108979.1 | | 54 | 107136.5 | 1.0172 |
| | $\zeta$(9d) | 5.0 | | fixed | 5.0 | 1.0000 |
| 5s$^2$10d | $E_{av}$ | 111111.2 | | 54 | 109247.0 | 1.0171 |
| | $\zeta$(10d) | 3.3 | | fixed | 3.3 | 1.0000 |
| 5s$^2$11d | $E_{av}$ | 112569.7 | 3 | 54 | 110689.9 | 1.0170 |
| | $\zeta$(11d) | 2.3 | | fixed | 2.3 | 1.0000 |
| 5s$^2$12d | $E_{av}$ | 113560.6 | 3 | 55 | 111722.9 | 1.0164 |
| | $\zeta$(12d) | 1.7 | | fixed | 1.7 | 1.0000 |
| 5s$^2$5g | $E_{av}$ | 100389.8 | | 54 | 98428.2 | 1.0199 |

| Configuration | Parameter | LSF | Group [a] | STD | HFR | LSF/HFR |
|---|---|---|---|---|---|---|
| | $\zeta(5g)$ | 0.1 | | fixed | 0.1 | 1.0000 |
| $5s^26g$ | $E_{av}$ | 105776.4 | | 54 | 103823.2 | 1.0188 |
| | $\zeta(6g)$ | 0.1 | | fixed | 0.1 | 1.0000 |
| $5s^27g$ | $E_{av}$ | 109026.5 | | 54 | 107087.6 | 1.0181 |
| | $\zeta(7g)$ | 0.0 | | fixed | 0.0 | |
| $5s^28g$ | $E_{av}$ | 111136.7 | | 54 | 109207.1 | 1.0177 |
| | $\zeta(8g)$ | 0.0 | | fixed | 0.0 | |
| $5s^29g$ | $E_{av}$ | 112585.0 | | 54 | 110660.2 | 1.0174 |
| | $\zeta(9g)$ | 0.0 | | fixed | 0.0 | |
| $5s^210g$ | $E_{av}$ | 113618.4 | | 54 | 111699.7 | 1.0172 |
| | $\zeta(10g)$ | 0.0 | | fixed | 0.0 | |
| $5s^211g$ | $E_{av}$ | 114383.9 | 4 | 54 | 112466.0 | 1.0171 |
| | $\zeta(11g)$ | 0.0 | | fixed | 0.0 | |
| $5s^212g$ | $E_{av}$ | 114944.3 | 4 | 55 | 113049.7 | 1.0168 |
| | $\zeta(12g)$ | 0.0 | | fixed | 0.0 | |
| 4f5s5p [c] | $E_{av}$ | 148823.6 | | fixed | 142153.7 | 1.0469 [c] |
| $5s5d^2$ [c] | $E_{av}$ | 196200.2 | | fixed | 189530.3 | 1.0352 [c] |
| | | | | | | |
| Configuration interaction [d] | | | | | | |
| $5s5p^2$-$5s^25d$ | $R^1$(5p5p,5s5d) | 18161.1 | 1 | 124 | 27501.0 | 0.6604 |
| $5s5p^2$-$5s^26d$ | $R^1$(5p5p,5s6d) | 9433.0 | 1 | 64 | 14284.3 | 0.6604 |
| $5s5p^2$-$5s^27d$ | $R^1$(5p5p,5s7d) | 6265.9 | 1 | 43 | 9488.4 | 0.6604 |
| $5s5p^2$-$5s^28d$ | $R^1$(5p,5p,5s,8d) | 4595.3 | 1 | 31 | 6958.6 | 0.6604 |
| $5s5p^2$-$5s^29d$ | $R^1$(5p5p,5s9d) | 3570.9 | 1 | 24 | 5407.4 | 0.6604 |
| $5s5p^2$-$5s^210d$ | $R^1$(5p5p,5s10d) | 2884.5 | 1 | 20 | 4368.0 | 0.6604 |
| $5s5p^2$-$5s^211d$ | $R^1$(5p5p,5s11d) | 2396.1 | 1 | 16 | 3628.4 | 0.6604 |
| $5s5p^2$-$5s^212d$ | $R^1$(5p5p,5s12d) | 2033.2 | 1 | 14 | 3078.9 | 0.6604 |

[a] Parameters in each numbered group were linked together with their ratio fixed at the Hartree-Fock level.

[b] All configuration-interaction parameters $R^k$ for the odd configurations were fixed at 80 % of the Hartree-Fock value.

[c] These highly excited configurations are unknown experimentally. They were included in the calculations in order to account for their interaction with other configurations studied in this work. Except for the average energies $E_{av}$ given here and $\zeta(5p)$ for $5p5d^2$ and $4f5p^2$, all other parameters of these configurations were fixed at the 80 % of the Hartree-Fock values ($F^k$, $G^k$, $R^k$) or 100 % of the Hartree-Fock values ($\zeta$).

[d] Other $R^k$ parameters of the even configurations were fixed at 80 % of the Hartree-Fock value.

Natural tin consists of ten stable isotopes with abundances ranging from 0.3 % to 33 %. Three of these isotopes have nuclear spin 1/2 and a rather large nuclear magnetic moment about $-1.0\,\mu_N$. Thus, lines observed from samples of natural tin (which were used in all experimental works quoted in the present paper) must be broadened by isotope shifts and hyperfine structure. Since there is no such entity as an atom of natural tin, the energy levels derived by our level optimization do not correspond to any physical object but are empirical values that best describe the observed spectral lines. This should be kept in mind when using the high-precision values from tables 1 and 3. Asymmetry of line profiles caused by isotope shifts and hyperfine structure may result in deviations of observed peak wavelengths from the Ritz values given in table 1. Observed isotope shifts between adjacent even isotopes are typically (0.005–0.02) $cm^{-1}$, while the hyperfine structure in less abundant odd isotopes is an order of magnitude larger. References to studies of isotope shifts and hyperfine structure of Sn II can be obtained from the NIST Atomic Energy Levels and Spectra Bibliographic Database at http://physics.nist.gov/Elevbib.

For completeness, we note that there is only one reported measurement of the Landé *g*-factor for Sn II. Namely, David et al. [27] accurately measured the Landé *g*-factor for the $5s5p^2\ ^4P_{3/2}$ level to be 2.6609(7).

### 4.4. Intensities of observed lines

In the history of atomic spectroscopy, it has been an unfortunate long-standing tradition to give very rough estimates of relative intensities of observed lines. Although line intensities were always recognized to be important in correct identification of transitions causing them, the arguments had to be qualitative because the sensitivity of registration strongly varies with wavelength and depends on rarely quantified properties of detectors, spectrographs, and optics used. Also, different excitation conditions in light sources lead to large variations in line intensities. A method suggested and successfully used in a recent series of papers [28–30] overcomes these problems and allows one to reduce line intensities observed by different authors using different equipment to a common uniform scale. The method is based on using the Boltzmann equation to approximate populations of energy levels together with theoretically estimated radiative rates. It was shown in the papers quoted above that this approximation in most cases allows one to describe the observed intensities by a simple formula with weighted transition rate (*gA*) multiplied by a Boltzmann factor with a suitable effective excitation temperature. Then spectral response functions of the registration equipment can easily be derived by comparing observed and modeled intensities, and intensities observed with different setups can be reduced to a uniform scale with a common excitation temperature, Deviations of plasma conditions from the local thermodynamic equilibrium (LTE) and inaccuracies in estimated transition rates and derived response functions of registration equipment typically lead to errors of about a factor of three in such modeled intensities. Nevertheless, thus derived intensities provide a robust quantitative criterion for line identification and can even be used to estimate transition rates, when such estimates cannot be obtained from theory. Of course, the above-mentioned factor-of-three uncertainty is a restriction for many applications, but there are many cases where such estimates can be useful.

This method was applied to obtain the reduced relative intensities given in table 1. Below, we explain reduction of intensities for each set of observations.

The Boltzmann plot for our observed line intensities, shown in figure 3a, indicates an effective excitation temperature of 2.0 eV in our triggered spark source. This plot was built with intensities corrected for the variation of response function of our equipment with wavelength, denoted as $I_{corr}$. The logarithmic intensity-correction function $F(\lambda)$ used for this correction is shown in figure 3b. Correction is made by multiplying the observed intensities by exponent of $F(\lambda)$. Transition rates *gA* used in the Boltzmann plots were calculated with Cowan's codes using our fitted parameters from the LSF.

Similarly, figures 3c and 3d present the Boltzmann plot and intensity-correction function for exposure 1 in Wu's line list [11]. It should be noted that the quantity given by Wu in the intensity columns is actually transparency (not the commonly used darkening) of the photographic plate on the scale 0 to 1000. To obtain the intensities, we subtracted his transparency values from 1000. Effective temperature in the source used for exposure 1 turned out to be 4.2 eV, which is the highest for all light sources used in the published literature. Apparently, this high temperature allowed Wu to observe lines from very highly excited levels not observed in other experiments. Reduction of intensities observed in the other three exposures reported by Wu [11] was made in a similar way. Effective temperatures for his exposures 2, 3, and 4 turned out to be about 3.6 eV, 3.7 eV, and 3.8 eV, respectively. Response functions derived from exposures 2 and 3, which cover the same wavelength range as exposure 1, are similar to the one shown in figure 3d. For the final reduction of Wu's intensity values, we used the correction function averaged over these three exposures.

It should be noted that, despite the non-linear properties of photographic plates, the original observed intensities in both our and Wu's work did not show any significant non-linearity with exposure. This can easily be verified by plotting the ratio of calculated and observed intensities versus the observed intensity. Non-linearity would result in a trend on such plots, which was not detected.

Intensities observed by Brill [10] and by McCormick and Sawyer [6] were reduced by the same method as described above. The effective excitation temperature in the light source used by Brill was found to be 1.9 eV, which is close to our triggered-spark value of 2.0 eV.

For the light source used by McCormick and Sawyer [6], the effective temperature was found to be somewhat lower, about 1.4 eV.

After the variations of response functions of registration equipment were removed from the observed intensities, and the effective temperatures were determined for each set of observations, it was easy to scale the corrected observed intensities to the same effective temperature. We chose the highest temperature in all sets of measurements, 4.2 eV, as the basis for the unified scale. This choice is motivated by the need to have the smallest range of final intensity values, which is convenient for presentation purposes.

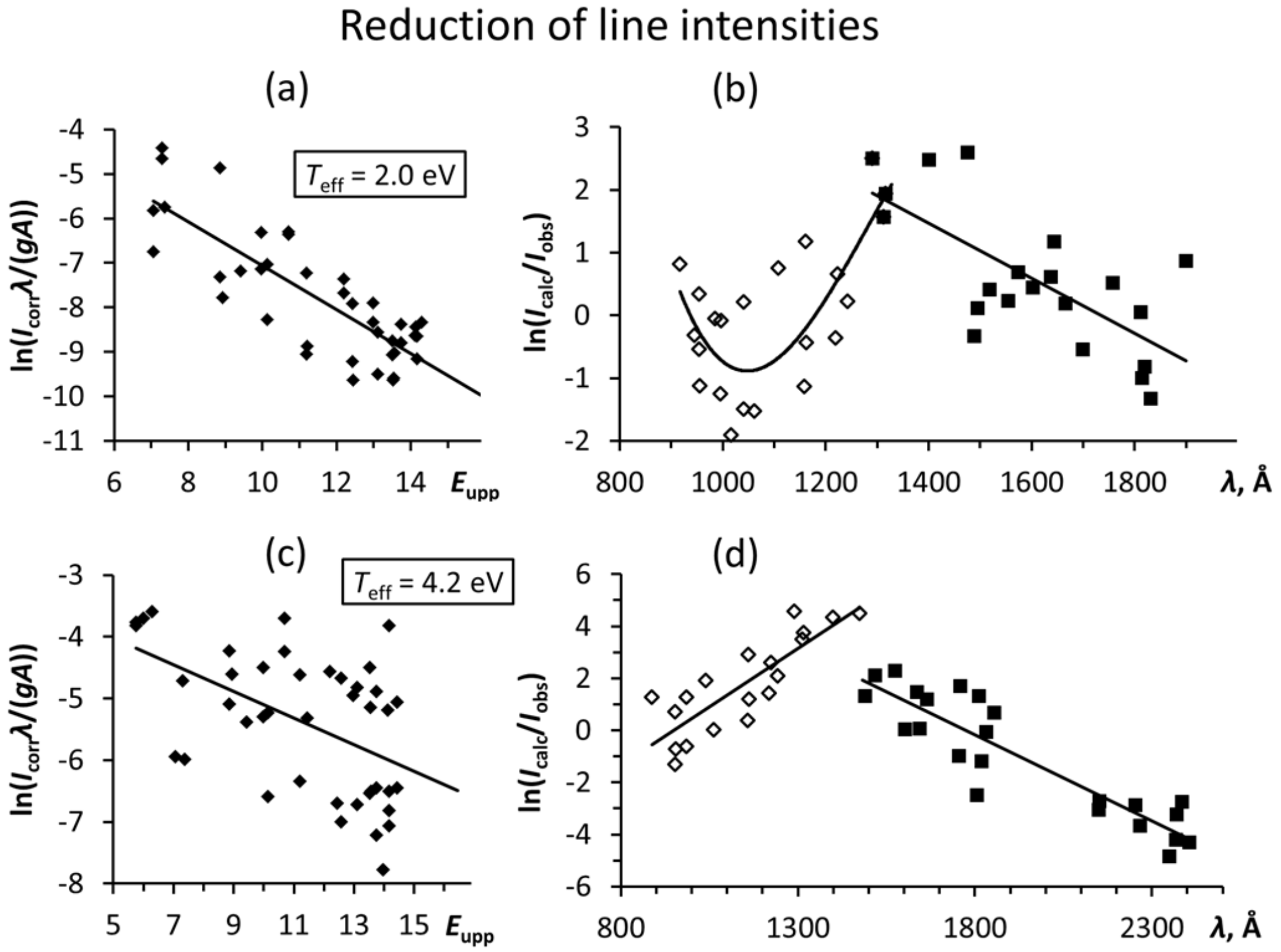

Figure 3. Boltzmann plots (a, c) and logarithmic intensity-correction functions (b, d) for our observations and those of Wu [11]. The upper-level energies $E_{upp}$ in the Boltzmann plots are given in eV. The effective temperatures derived from the negative slope of the Boltzmann plots are shown in boxes. The calculated intensities $I_{calc}$ in panels b and d are obtained from weighted transition rates $gA$ calculated in the present work with a formula $I_{calc} = (gA/\lambda)\exp(-E_{upp}/T_{eff})$.

### *4.5. Ionization potential*

The ionization potential (IP) given in AEL [4] is the value obtained by McCormick and Sawyer [6] using the $5s^2ng$ ($n$ = 6–11) series. As the 5g level was established in the present work, and the measurements of McCormick and Sawyer contained significant systematic shifts, the IP has to be revised. We obtained the new value of IP using both the Ritz-type quantum-defect series extrapolation and core-polarization formula fitting for the $5s^2ng$ $^2G$ ($n$ = 5–11) series using computer codes RITZPL and POLAR [31]; both leading to almost the same value. The formulas used in these series-fitting computer codes and explanation of their application can be found, for example, in reference [28]. The IP obtained from RITZPL using the two-parameter extended Ritz formula was 118023.6(7) $cm^{-1}$ and that from POLAR was 118023.8(7) $cm^{-1}$. Fitting of the three-parameter extended Ritz formula for the $5s^2ns$ ($n$ = 6–11) series yields 118036.4(2) $cm^{-1}$ for the ionization energy. It is known that the $n$s series is slightly perturbed by an interaction with $5s5p^2$ $^2S_{1/2}$. The $n$g series is free from such perturbations. Therefore, we adopted the average IP value obtained from the two fits of the $5s^2ng$ series, 118023.7(5) $cm^{-1}$, which is equivalent to 14.63307(6) eV. All fits were made using weights inversely proportional to squared uncertainties of the level values from table 2 combined in quadrature with the uncertainty of the ground level, 0.04 $cm^{-1}$. Our value is 6.7 $cm^{-1}$ higher than the previously recommended value from McCormick and Sawyer [6].

## 5. Comparison with observed Auger electron spectrum

The Auger electron spectrum of Sn I and Sn II in the low-energy region 0–20 eV was observed by Forrest et al. [32] in a crossed atomic and electron beams experiment. They assigned several observed peaks to autoionization decay of the $4d^95s^25p^2$ configuration of Sn II, not considered in this work. In addition, they tentatively assigned a strong peak observed at 2.529 eV to the autoionization decay of the $5p^3$ $^2P°$ term of Sn II. This assignment does not agree with our identifications. According to our parametric fitting, the $5p^3$ configuration is highly mixed with 5s5p5d, and the largest contribution of $5p^3$ $^2P°$ is predicted for the levels with large contributions from $5s5p(^3P°)5d$ $^2P°$ at about 128000 $cm^{-1}$ and $5s5p(^1P°)5d$ $^2P°$ at about 152000 $cm^{-1}$. Autoionization decay of these levels to the $5s^2$ ground state of Sn III would produce Auger peaks at about 1.2 eV and 4.3 eV, respectively. Forrest et al. [32] observed a weak peak at 1.023 eV and medium-strength peaks at 4.117 eV and 4.277 eV, which may be associated with these predicted levels. However, for the peak at 1.023 eV our calculations yield a higher autoionization rate from a close predicted $5s5p(^3P°)5d$ $^2F°_{5/2}$ level at about 127000 $cm^{-1}$.

A few of the peaks observed by Forrest et al. [32] closely match the experimental energies of autoionizing Sn II levels we derived from our observed optical spectrum. In particular, the peaks observed at 1.761 eV and 1.829 eV closely match the predicted Auger energies for the $5s5p(^1P°)5d$ $^2D°$ $J = 3/2$ and 5/2 levels (observed at 132168.95 $cm^{-1}$ and 132708.1 $cm^{-1}$), respectively.

The peak observed at 0.657 eV can be a blend of Auger decays of the $5p^3$ $^4S°_{3/2}$ and $5s5p(^3P°)5d$ $^4P°_{5/2}$ levels (which we observed at 123156.8 $cm^{-1}$ and 123688.3 $cm^{-1}$, respectively). These decays are predicted to be of comparable strengths, due to small admixtures of doublet terms in the composition of these levels.

The peak at 0.285 eV was assigned by Forrest et al. [32] to the decay of the Sn I $5s5p^3$ $^3P°_1$ and level to the $5s^25p$ $^2P°_{1/2}$ ground level of Sn II. However, this assignment was later rejected by Dembczynski and Wilson [33]. This peak closely matches our observed energy for the Sn II $5s5p(^3P°)5d$ $^4D°_{5/2}$ level (120253.85 $cm^{-1}$, corresponding to the Auger electron energy of 0.2773 eV), while the observed peak at 0.523 eV matches the decay of the $5s5p(^3P°)5d$ $^4D°_{7/2}$ level (122491.6 $cm^{-1}$, corresponding to the Auger electron energy of 0.5548 eV).

Finally, our calculations predict the metastable $5s5p(^3P°)5d$ $^4F°_{9/2}$ level at 118700 $cm^{-1}$. Autoionization of this level should produce an Auger peak at ejected electron energy of 0.088 eV. The strongest peak observed by Forrest et al. [32] is at 0.053 eV. This peak may be due to the decay of this metastable level.

We note that autoionization rates calculated for the Sn II levels discussed in this section are unreliable, because they strongly depend on very small mixing between doublet and quartet levels and on poorly known interaction between the $5p^3$ and 5s5p5d configurations. This, as well as the low resolution of the observed Auger electron spectrum [32], precludes definite identification of the observed Auger features. More sophisticated calculations, as well as higher-resolution experiments, are needed to elucidate the structure of autoionizing Sn II levels in the region just above the first ionization limit.

## 6. Transition probabilities

Oliver and Hibbert [14] made a large-scale Breit-Pauli configuration-interaction (CI) calculation of transition probabilities of Sn II using the CIV3 code of Hibbert and co-workers (see references in [14]). They presented three sets of results: one for their *ab initio* calculation (in the length gauge) and two for the fine-tuned calculation (one in the length gauge and the other in the velocity gauge). The fine tuning consisted of semiempirical adjustment of the diagonal matrix elements of the Hamiltonian minimizing the differences between the calculated and experimental eigenvalues. The line strengths $S_L$ obtained in the length gauge in the fine-tuned calculation are considered to be the most accurate ones from the three sets. Their accuracy can be assessed by comparing them with the other two data sets, $S_v$ (fine-tuned, velocity gauge) and $S_{ab}$ (*ab initio*, length gauge). This comparison, illustrated in figure 4, shows that for strong lines with $S_L > 0.28$ the length and velocity forms of line strength agree within 6 % on average, while for weaker lines with $S_L$ = (0.03–0.28) the agreement is somewhat worse, about 12 % on average. We adopted these standard deviations as conservative estimates of uncertainties of $S_L$ in the corresponding ranges of line strength.

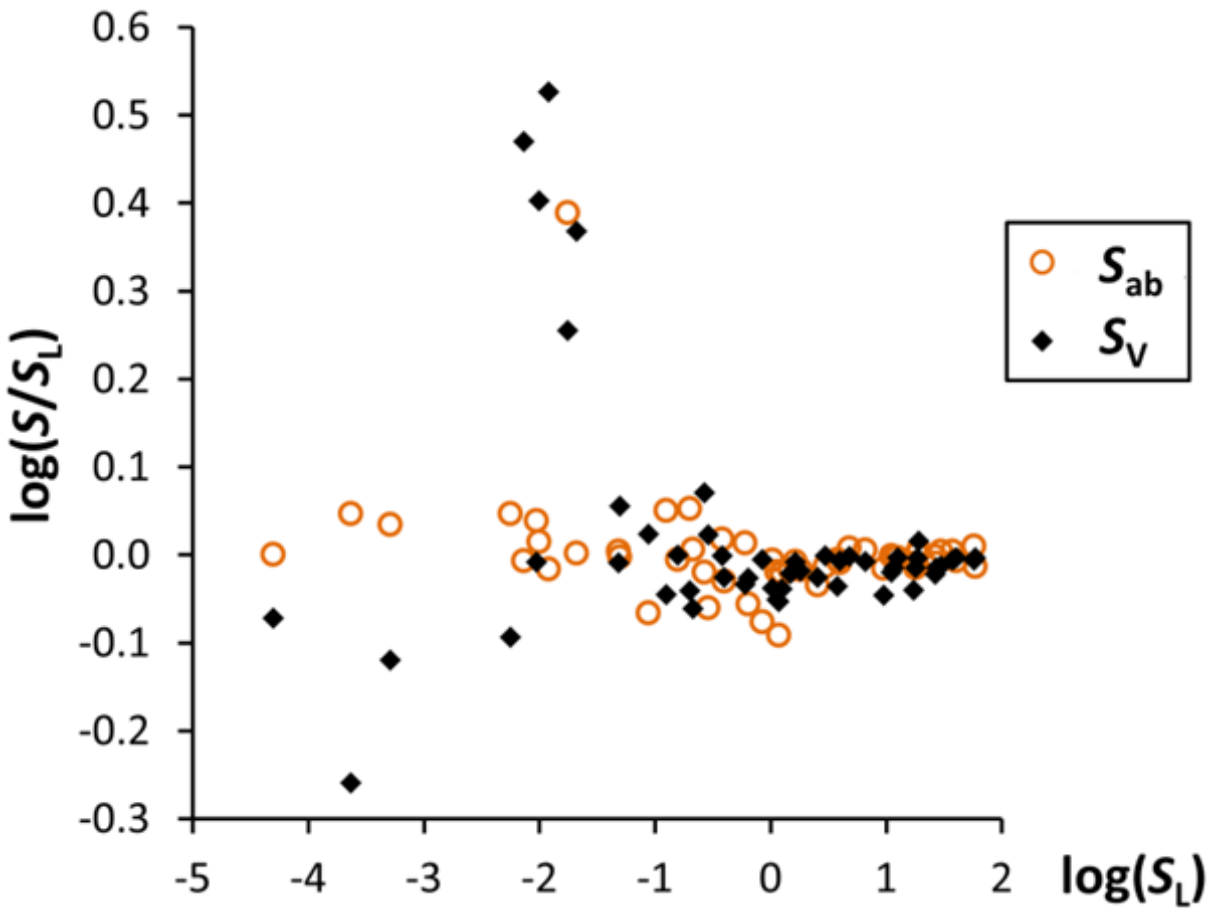

Figure 4. (Color online) Comparison of line strengths $S$ calculated by Oliver and Hibbert [14] in different approximations and gauges: $S_L$ – fine-tuned calculation in length gauge; $S_V$ – fine-tuned calculation in velocity gauge; $S_{ab}$ – *ab initio* calculation in length gauge.

For the ten weakest lines with $S_L < 0.03$, the length and velocity forms strongly disagree with each other. Most of these transitions are intercombination ones between doublet and quartet levels. As pointed out by Oliver and Hibbert [14], for such transitions, calculation of the line strength in velocity gauge requires additional terms not accounted for in the CIV3 code. This makes the comparison meaningless for intercombination transitions. Instead, the comparison of the *ab initio* and fine-tuned calculations in the length gauge can be used for estimating their uncertainties. Except for one large deviation for the $5s^25p$ $^2P°_{3/2}$ – $5s5p^2$ $^2S_{1/2}$ transition at 1394.667 Å, $S_{ab}$ agrees with $S_L$ within 12 %. However, because of low statistics, we adopt a conservative estimate of 35 % for the uncertainty of transitions with $S_L$ = (0.001–0.03) and omit the three weakest transitions, for which the transition rate given by Oliver and Hibbert [14] strongly contradicts the observed line intensities.

The high accuracy of calculations of Oliver and Hibbert [14] for strong lines is further confirmed by comparison of calculated and observed radiative lifetimes presented in table 5.

David et al. [27], employing the direct magnetic resonance method, measured the lifetime of the $5s5p^2$ $^4P_{1/2}$ level in Sn II to be 325(40) ns. They supported this result by two additional less accurate measurements with two independent methods.

Schectman et al. [15] measured the lifetimes of three levels, $5s^25d$ $^2D_{3/2,5/2}$ and $5s^24f$ $^2F°_{5/2}$, with a beam-foil method. Using a similar method, Andersen and Lindgård [16] measured the lifetime of the $5s^26s$ $^2S_{1/2}$ and $5s^25d$ $^2D_{3/2}$ levels. Both these studies carefully accounted effects of cascades on the measured decay curves.

Table 5. Comparison of observed and calculated lifetimes in Sn II

| Level | | Energy $cm^{-1}$ | $\tau_{obs}$ ns | Ref.[a] | $\tau_{th}$ ns | Ref.[a] |
|---|---|---|---|---|---|---|
| $5s5p^2$ | $^4P_{1/2}$ | 46464.301 | **325(40)** | D80 | **375** | OH10 |
| | | | 1500[b] | AM00 | 215 | TW |
| | | | | | 237 | AM05 |
| $5s^26s$ | $^2S_{1/2}$ | 56886.3763 | **1.10(10)** | AL77 | **1.16** | OH10 |
| | | | | | 1.20 | TW |
| | | | | | 1.13 | AM05 |
| $5s^25d$ | $^2D_{3/2}$ | 71406.155 | **0.44(2)** | S00 | **0.45** | OH10 |
| | | | 0.50(5) | AL77 | 0.37 | TW |
| | | | | | 0.41 | AM05 |
| $5s^25d$ | $^2D_{5/2}$ | 72048.273 | **0.46(4)** | S00 | **0.51** | OH10 |
| | | | | | 0.45 | TW |
| | | | | | 0.50 | AM05 |
| $5s^24f$ | $^2F°_{7/2}$ | 89288.268 | **5.0(10)**[c] | GV85 | **3.82** | OH10 |
| | | | 6.9[b] | AM00 | 3.28 | TW |
| | | | | | 3.21 | AM05 |
| $5s^24f$ | $^2F°_{5/2}$ | 89294.068 | **4.6(10)** | S00 | **3.78** | OH10 |
| | | | 5.2(10)[c] | GV85 | 3.24 | TW |
| | | | 4.8[b] | AM00 | 3.04 | AM05 |

[a] References: AL77 – Andersen and Lindgård [16]; AM00 – Alonso-Medina and Colón [34]; AM05 – Alonso-Medina et al. [36] (Cowan code); D80 – David et al. [27]; GV85 – Gorshkov and Verolainen [37]; OH10 – Oliver and Hibbert [14]; S00 – Schectman et al. [15]; TW – this work (Cowan code).
[b] Determined from the sum of measured radiative rates.
[c] Original estimate of uncertainty doubled (see text).

Gorshkov and Verolainen [37] determined the lifetimes of the two $5s^24f$ $^2F°_{5/2,7/2}$ levels by using intersecting atomic and electron beams and a multichannel method of retarded coincidences. Although they reported very small uncertainties of ±0.5 ns, their description of the experiment lacks any mention of an account for cascading effects. Therefore, in table 5 we have doubled their uncertainty estimate.

As can be seen from table 5, lifetimes calculated by Oliver and Hibbert [14] agree with all the best measurements within the uncertainties.

Our own calculations made with the Cowan codes (using the LSF parameters) are compared with the calculations of Oliver and Hibbert [14] in figure 5.

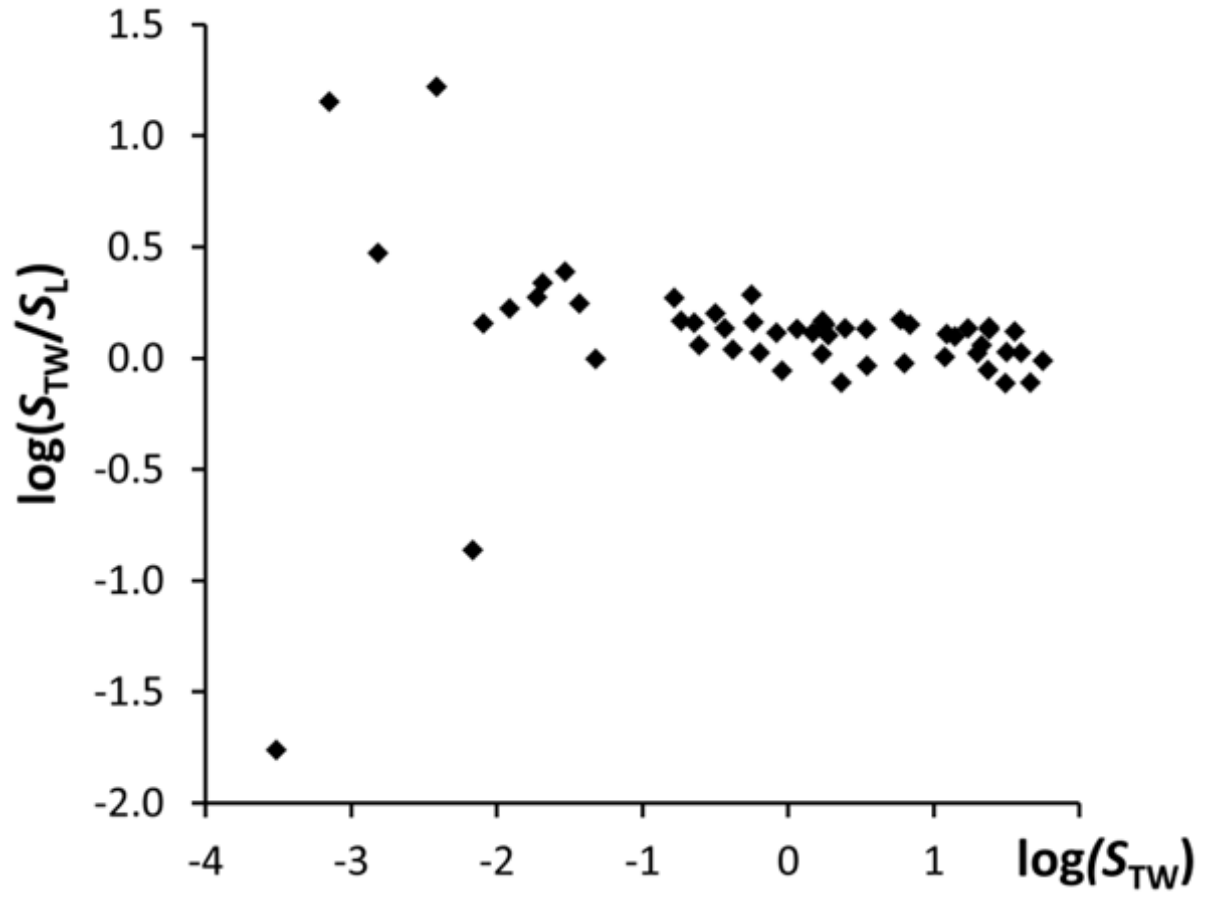

Figure 5. Comparison of line strengths calculated in the present work with Cowan's codes ($S_{TW}$) with those from fine-tuned calculations of Oliver and Hibbert in the length gauge ($S_L$).

For strong transitions with line strength $S > 0.5$, our calculations agree with those of Oliver and Hibbert [14] to 28 % on average. For weaker transitions, the results of Cowan's codes deviate from Oliver and Hibbert [14] by more than a factor of two on average. Calculations of Alonso-Medina et al. [36], also using a parametric fitting with Cowan's codes, are of similar quality, although they display somewhat larger deviations from Oliver and Hibbert [14] (about 30 % on average for $S > 1$, and 70 % for weaker transitions). We note that the *f*- and *A*-values given by Alonso-Medina et al. [36] in their table V for the $5s^2nf–5s^2n'g$ transitions are not consistent with each other and strongly disagree with our calculations.

Results of Oliver and Hibbert [14] also compare well with the relativistic all-order calculations of Safronova et al. [38]. These authors presented their results only for a few $5s^2ns–5s^2n'p$ and $5s^2np–5s^2n's$ transitions. They agree with Oliver and Hibbert [14] with an average deviation of 12 %, except for one $5s^25p\ ^2P°_{3/2} – 5s^27s\ ^2S_{1/2}$ transition (1219.083 Å), for which their *S* value is lower by a factor of 2.5.

Aside from a few discrepancies mentioned above, theoretical calculations of line strengths agree with each other, at least for strong transitions, and they agree reasonably well with the few available lifetime measurements. However, comparison with experimentally measured radiative rates (*A*-values) presents problems. The *A*-values were measured for several tens of transitions by Alonso-Medina and Colón [34], Schectman et al. [15], Miller et al. [39], Wujec and Weniger [40], and Wujec and Musielok [41]. Experimental line strengths reported in these papers are compared with the critically evaluated theoretical data in figure 6. Only a few measured values agree with theory within the claimed measurement uncertainties. The greatest discrepancies are observed for the weakest lines measured by Alonso-Medina and Colón [34]. It is difficult to identify the causes of the discrepancies. However, from the above analysis of the theoretical data, we conclude that the discrepancies originate in some flaws in the measurements. For this reason, we retained in table 1 only four experimental *A*-values, three from Alonso-Medina and Colón [34] and one from Miller et al. [39], and assigned greatly increased uncertainties to them.

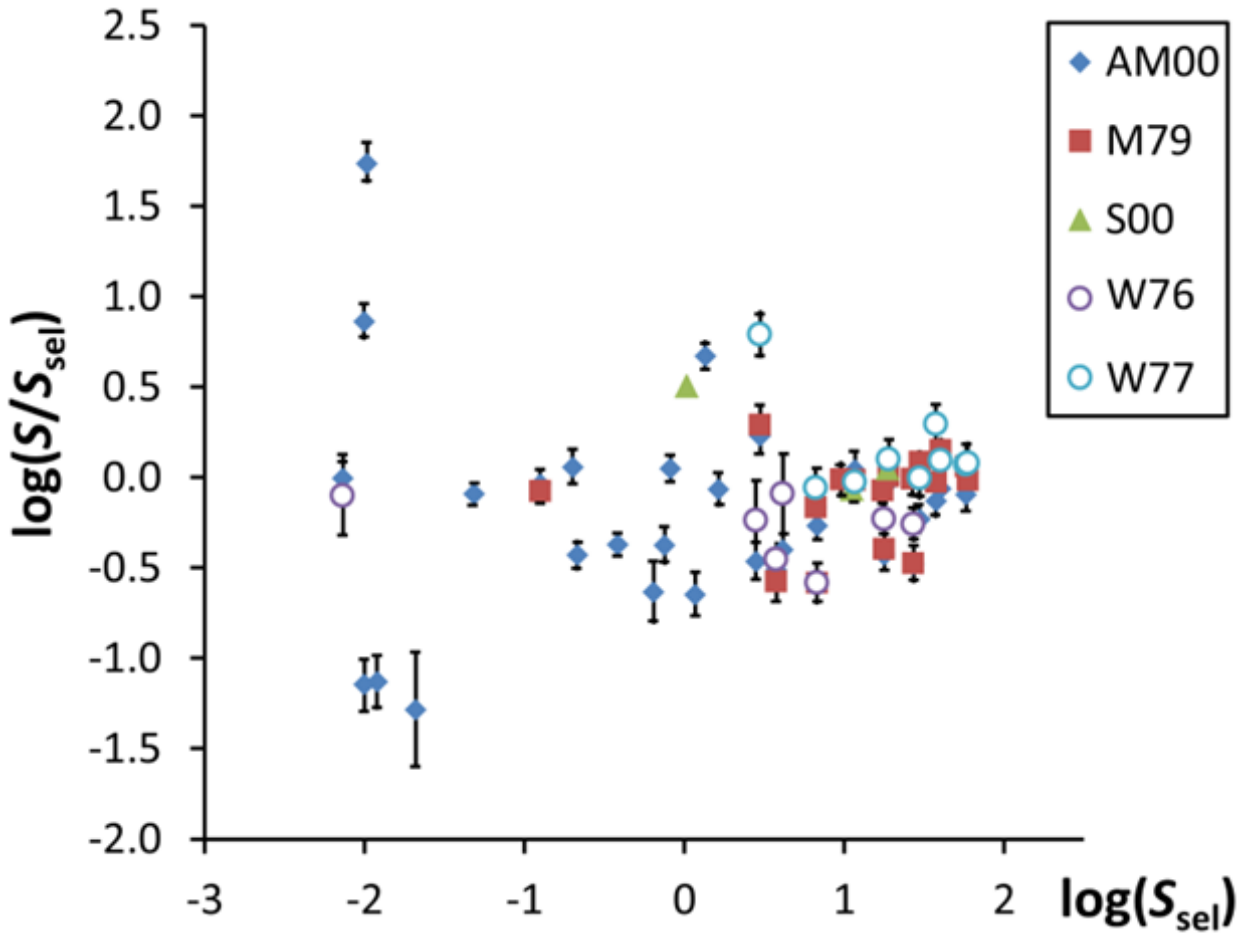

Figure 6. (Color online) Comparison of experimental line strengths *S* with selected theoretical data. The selected line strength $S_{sel}$ were taken from Oliver and Hibbert [14] and from our calculations and have estimated uncertainties between 6 % and 35 %. The error bars correspond to claimed measurement uncertainties (one standard deviation). Key to experimental work: AM00 – Alonso-Medina and Colón [34]; M79 – Miller et al. [39]; S00 – Schectman et al. [15]; W76 – Wujec and Musielok [41]; W77 – Wujec and Weniger [40].

We included in table 1 four lines at 2442.7 Å, 2486.6 Å, 2592.3 Å, and 3351.3 Å, for which Alonso-Medina and Colón [34] reported measured *A*-values. Since these authors did not attempt to accurately measure the wavelengths, and these lines were not reported by other authors, the wavelengths given in the column $\lambda_{obs}$ are actually the rounded Ritz wavelengths. We note that the last two of these lines, as well as two other lines reported by Alonso-Medina and Colón [34] at 2592.6 Å and 3351.9 Å, were incorrectly identified by these authors.

We also included in table 1 one unobserved parity-forbidden line corresponding to the transition between the levels of the ground term. Our predicted wavelength for this far-infrared line is 23521.08(23) Å.

According to calculations of Biémont et al. [35], Warner [42], and Garstang [43], this line is dominated by the magnetic dipole (M1) transition. The *A*-values calculated for this M1 transition in these works agree with each other within 1 %. The *A*-value for the electric quadrupole transition, 2.893 $s^{-1}$ [35], amounts to only 0.4 % of the M1 decay rate and can be neglected in most applications.

Since the statistical distribution of both measured and calculated *A*-values is far from normal, uncertainties of the adopted *A*-values are specified in table 1 with a letter code instead of numerical values. The letter code is explained in table 6.

Table 6. Transition probability uncertainty code

| Letter | Uncertainty in *A*-value | Uncertainty in log(*gf*) |
|---|---|---|
| AAA | ≤ 0.3 % | ≤ 0.0013 |
| AA | ≤ 1 % | ≤ 0.004 |
| A+ | ≤ 2 % | ≤ 0.009 |
| A | ≤ 3 % | ≤ 0.013 |
| B+ | ≤ 7 % | ≤ 0.03 |
| C+ | ≤ 18 % | ≤ 0.08 |
| C | ≤ 25 % | ≤ 0.11 |
| D+ | ≤ 40 % | ≤ 0.18 |
| D | ≤ 50 % | ≤ 0.24 |
| E | > 50 % | > 0.24 |

## 7. Conclusion

A comprehensive interpretation of the spectrum of singly ionized tin (Sn II) is presented here. The analysis covers the wavelength region 887 Å to 10611 Å. The earlier reported levels of even parity configurations, $5s^2nd$ ($n$ = 5–11), $5s^2ns$ ($n$ = 6–11), $5s^2ng$ ($n$ = 6–11) and $5s5p^2$ have been confirmed with minor improvements in their level values, while the $5s^25g$ level has been newly identified. The ambiguity in the level values of $^2S_{1/2}$ and $^2P_{1/2}$ of the $5s5p^2$ configuration has been resolved. In odd parity, the reported levels of the $5s^2np$ ($n$ = 5−9) and $5s^2nf$ ($n$ = 4−6) configurations have been verified. Sixty-nine levels are now known in Sn II. Among these, eight are new, and for 11 levels previous values and/or interpretations have been revised. The level values, which are based on the identification of about 200 spectral lines, have been optimized in a least-squares fitting procedure. About 70 of these lines were measured by us either for the first time or with a significantly improved precision. With these improved data, the ionization energy of Sn II has been determined more accurately. For 140 transitions out of total 215, we give a critically evaluated value of transition probability with an estimated uncertainty. About 40 % of these transition probabilities have an accuracy C+ (≤ 18 %) or better.

**Acknowledgements**

KH would like to duly acknowledge the Council of Scientific and Industrial Research (CSIR) for giving financial support through the Senior Research Fellowship (SRF) Scheme. AT is thankful to the kind hospitality of late Prof. Y N Joshi and St. Francis Xavier University, Antigonish (Canada) during the recording of tin spectra.